\documentclass[aps,prd,reprint,showpacs,nofootinbib,showkeys,superscriptaddress,floatfix]{revtex4-1}
\usepackage{empheq}

\pdfoutput=1  
\usepackage{soul}
\usepackage{amsmath}
\usepackage{amssymb}
\usepackage{tikz}
\usepackage{multirow}
\usepackage[american]{babel}
\usepackage{booktabs}
\usepackage{cancel}
\usepackage{dcolumn}  
\usepackage[T1]{fontenc}  
\usepackage{graphicx}  
\usepackage[]{hyperref}  
\usepackage[utf8]{inputenc}  
\usepackage{url}
\usepackage[dvipsnames]{xcolor}
\usepackage{float}
\usepackage[normalem]{ulem}
\usepackage{textcase}
\usepackage{subcaption}
\usepackage{dashrule}
\usepackage{booktabs}
\usepackage{multirow}
\usepackage[table]{xcolor}

\usepackage[table,dvipsnames]{xcolor}

\DeclareUnicodeCharacter{2212}{-}

\usepackage{tikz}
\usepackage{array}

\usetikzlibrary{arrows,shapes,positioning,shadows,trees}

\tikzset{
	basic/.style  = {draw, text width=2cm, drop shadow, font=\sffamily, rectangle},
	root/.style   = {basic, rounded corners=2pt, thin, align=center,
		fill=green!30},
	level 2/.style = {basic, rounded corners=6pt, thin,align=center, fill=green!60,
		text width=8em},
	level 3/.style = {basic, thin, align=left, fill=pink!60, text width=6.5em}
}

\usetikzlibrary{arrows,decorations.pathmorphing,backgrounds,fit,positioning,shapes.symbols,chains}

\newcolumntype{d}[1]{D{.}{.}{#1}}
\newcolumntype{v}[1]{D{,}{,\ }{#1}}

\makeatletter

\newcommand{\Rmnum}[1]{\expandafter\@slowromancap\romannumeral #1@}

\newcommand*\circled[1]{\tikz[baseline=(char.base)]{\node[shape=circle,draw,inner sep=1pt] (char) {#1};}}

\makeatother

\renewcommand {\arraystretch}{1.3}

\usepackage{hyperref}
\hypersetup{
    colorlinks=true, 
    linkcolor=blue, 
    citecolor=magenta}

\begin{document}

\title{Beyond $\Lambda$CDM: Probing Structure Growth and Observational Viability of Generalized ${f\,(Q , \mathcal{L}_{m})}$ Gravity}

\author{Raj Kumar Das}
\email{raj1996cool@gmail.com}
\affiliation{Physics and Applied Mathematics Unit, Indian Statistical Institute, 203 B. T. Road, Kolkata 700108, India}

\author{Sivasish Paul}
\email{sivasishpaul@mails.ucas.ac.cn}
\affiliation{School of Physical Sciences, University of Chinese Academy of Sciences, Beijing 100049, China}


\begin{abstract}
$f(Q, \mathcal{L}_m)$ gravity is a well-known extension of $f(Q)$ gravity, obtained through non-minimal or additive minimal coupling in the action's Lagrangian density. Background-level cosmography within this framework, along with comparisons to $\Lambda$CDM, has been the focus of prior studies. The present work considers a generalized power-law form of the additive minimal coupling, $f(Q,\mathcal{L}_m) = -Q^\mu + \alpha \mathcal{L}_m^{\,\nu} + \beta$, from which the Hubble function $H(z)$ is derived through a novel parametrization. When $\nu \neq 1$, the matter continuity equation is rendered non-conserved, a feature that can be interpreted as an effective particle-creation mechanism. Constraints on the model parameters are obtained from Observational Hubble Data, DESI DR2 BAO measurements, and three SNe Ia compilations (Pantheon+, Union3, DES-SN5YR), allowing their cosmographic effects to be examined. Beyond the background level, the perturbed field, Euler, and continuity equations are derived under non-conservation, and their impact on structure growth is studied through a joint analysis incorporating the $f\sigma_8$ probe. Bayesian evidence, computed relative to $\Lambda$CDM, shows that the model is favoured by background data in several cases, with moderate support for one realisation, whereas the inclusion of $f\sigma_8$ drives $\nu\to1$ and decisively disfavours the additional freedom, highlighting the importance of growth data in testing such extensions.
\end{abstract}

\maketitle
\section{Introduction}
\label{sec1}

The serendipitous discovery of the accelerated expansion of the Universe in the late 1990s~\cite{SupernovaSearchTeam:1998fmf, SupernovaCosmologyProject:1998vns} has profoundly reshaped our understanding of cosmic history and has prompted a rigorous scrutiny of some of the most fundamental questions in cosmology. In light of these late-time observations, the $\Lambda$CDM model, formulated within General Relativity (GR), has emerged as the concordance framework for describing cosmic evolution. Characterised by a cosmological constant $\Lambda$ driving the accelerated expansion together with cold dark matter (CDM), this model has demonstrated remarkable consistency with a wide range of astronomical observations. However, despite its many successes, both GR and the $\Lambda$CDM paradigm continue to face persistent challenges, including the cosmological constant problem~\cite{Weinberg:1988cp}, the cosmic coincidence problem~\cite{Zlatev:1998tr}, and the cosmological tensions that have emerged in recent observations~\cite{DiValentino:2021izs,Perivolaropoulos:2021jda,Abdalla:2022yfr,Kamionkowski:2022pkx,CosmoVerse:2025txj}. Moreover, the absence of a fundamental explanation for the dark sector, namely dark matter and especially dark energy, whose effects dominate the late-time dynamics of the Universe~\cite{Ishak:2018his, Joyce:2014kja}, further highlights the limitations of the standard paradigm. These unresolved issues have stimulated the exploration of cosmological scenarios beyond $\Lambda$CDM, either within GR or through extensions of the underlying theory of gravity, and a wide range of modified gravity theories have consequently been proposed and extensively studied~\cite{DeFelice:2010aj, Sotiriou:2008rp,Nojiri:2010wj, Nojiri:2017ncd, Capozziello:2011et, Clifton:2011jh, Bueno:2016dol, Khosravi:2016kfb, Haghani:2015zga, Quiros:2019ktw, Mishra:2020saj}.

Among these, an important direction extends the geometric foundations of gravitation beyond the Riemannian description underlying GR, by formulating gravity in terms of alternative geometric quantities such as torsion and non-metricity. Early developments along these lines can be traced back to the pioneering work of Hermann Weyl, who introduced the concept of non-metricity through a non-vanishing covariant derivative of the metric tensor, although his original attempt to unify gravitation and electromagnetism was ultimately unsuccessful~\cite{ORaifeartaigh:2000dtc,Heisenberg:2018vsk}. Nevertheless, these ideas laid the groundwork for subsequent geometric reformulations of gravity. For instance, the teleparallel equivalent of General Relativity (TEGR) describes gravity through the Weitzenb\"{o}ck connection, which has vanishing curvature and non-metricity but non-zero torsion~\cite{Cai:2015emx,Chen:2010va,Bahamonde:2021gfp,Dent:2010nbw,Wu:2010mn}. Complementarily, symmetric teleparallel gravity (STG) adopts an affine connection for which both curvature and torsion vanish, so that the gravitational interaction is entirely encoded in the non-metricity tensor $Q_{\alpha\mu\nu} = \nabla_{\alpha}g_{\mu\nu}$, where $\nabla_{\alpha}$ is the covariant derivative associated with this connection~\cite{BeltranJimenez:2017tkd,Heisenberg:2023lru, BeltranJimenez:2019tme}.

Building upon this framework, symmetric teleparallel gravity naturally admits extensions in which the gravitational action is generalised beyond its linear dependence on the non-metricity scalar $Q$. A prominent example is $f(Q)$ gravity, in which the gravitational Lagrangian is an arbitrary function of $Q$~\cite{BeltranJimenez:2017tkd,BeltranJimenez:2019tme}. In its linear limit the theory reproduces GR in the coincident gauge, up to a boundary term, while offering a novel geometric interpretation of the gravitational interaction. Owing to this structure, $f(Q)$ gravity has been extensively explored in cosmology, particularly in relation to the late-time accelerated expansion of the Universe~\cite{BeltranJimenez:2019tme,Heisenberg:2023lru}.

Further generalisations arise through couplings between geometry and matter. In this direction, $f(Q,T)$ gravity, where $T$ denotes the trace of the energy-momentum tensor, has been widely investigated and shown to exhibit a rich phenomenology for both the cosmic expansion and the growth of large-scale structure~\cite{Xu:2019sbp,Najera:2021afa}. A more general extension is provided by $f(Q,\mathcal{L}_m)$ gravity, in which the gravitational action depends explicitly on both the non-metricity scalar $Q$ and the matter Lagrangian density $\mathcal{L}_m$~\cite{Myrzakulov_2023,Myrzakulov:2024fQLm,Myrzakulov:2025ycj,Myrzakulov:2024lateQLm, Hazarika:2024fQLm}. In its general form this framework allows for a non-minimal coupling between geometry and matter, which modifies the standard conservation law of the energy-momentum tensor and enables an effective exchange of energy between the gravitational and matter sectors. Such features can significantly alter both the background evolution and the growth of matter perturbations, making this framework particularly suitable for testing deviations from $\Lambda$CDM.

In this work we consider a generalised additive form of $f(Q,\mathcal{L}_m)$ gravity, in which geometry and matter are minimally coupled, i.e.\ the action contains no cross terms between $Q$ and $\mathcal{L}_m$. Even in this case, a non-linear dependence on the matter Lagrangian prevents the matter sector from being separately conserved, so that both the expansion history and the growth of structure depart from their $\Lambda$CDM counterparts. We confront this scenario with cosmic chronometers, DESI BAO and three Type Ia supernova compilations (Pantheon+, Union3 and DESY5), both with and without redshift-space distortion measurements of $f\sigma_8$. We derive the evolution of matter perturbations in this framework, constrain the model parameters through MCMC analyses, and assess the statistical preference relative to $\Lambda$CDM through the Bayesian evidence.

\section{\boldmath ${f\,(Q\, , \mathcal{L}_{m})}$ Gravity}
\label{sec:mod_theory}

The $f(Q,\mathcal{L}_m)$ theory of gravity is formulated within the framework of non-Riemannian $f(Q)$ geometry, where gravity is described by the non-metricity scalar $Q$. This geometric construction itself represents a departure from the standard Riemannian description of gravity. The gravitational action is further generalized by allowing an explicit dependence on the matter Lagrangian $\mathcal{L}_m$, whose variation contributes directly to the gravitational dynamics and leads to modified field equations.

\begin{equation}
    S=\int f(Q,\mathcal{L}_m)\sqrt{-g}\, d^4x .
\end{equation}

The non-metricity scalar $Q$ originates from the non-metricity tensor defined as
\begin{equation}
    Q_{\alpha\mu\nu} = \nabla_\alpha g_{\mu\nu}
    = \partial_\alpha g_{\mu\nu}
    - Y^\beta_{\;\;\alpha\mu} g_{\beta\nu}
    - Y^\beta_{\;\;\alpha\nu} g_{\mu\beta} ,
\end{equation}
where $Y^{\alpha}{}_{\mu\nu}$ is the affine connection related to STEGR, which modifies the standard Levi--Civita connection by including contortion $K^{\alpha}{}_{\mu\nu}$ and disformation $L^{\alpha}{}_{\mu\nu}$ tensors. Using these definitions, the non-metricity scalar can be constructed as
\begin{equation}
    Q = - Q_{\alpha\mu\nu} P^{\alpha\mu\nu} .
\end{equation}

Here, $P^{\alpha\mu\nu}$ denotes the superpotential tensor, which is defined as
\begin{equation}
    P^{\alpha}{}_{\mu\nu}
    = -\frac{1}{2} L^{\alpha}{}_{\mu\nu}
    + \frac{1}{4} \left( Q^{\alpha} - \tilde{Q}^{\alpha} \right) g_{\mu\nu}
    - \frac{1}{4} \delta^{\alpha}_{(\mu} Q_{\nu)} .
\end{equation}

The disformation tensor is given by
\begin{equation}
    L^{\alpha}{}_{\mu\nu}
    = \frac{1}{2} Q^{\alpha}{}_{\mu\nu}
    - Q_{(\mu}{}^{\alpha}{}_{\nu)} .
\end{equation}

Varying the action with respect to the metric, we obtain
\begin{equation}
    \delta S = \int \Bigl[
    \left( f_Q\, \delta Q + f_{\mathcal{L}_m}\, \delta \mathcal{L}_m \right) \sqrt{-g}
    + f\, \delta \sqrt{-g}
    \Bigr] d^4x .
\end{equation}

Here,
\[
f_Q = \frac{\partial f}{\partial Q}, \quad
f_{QQ} = \frac{\partial^2 f}{\partial Q^2}, \quad
f_{\mathcal{L}_m} = \frac{\partial f}{\partial \mathcal{L}_m}, \quad
f_{\mathcal{L}_m \mathcal{L}_m}
= \frac{\partial^2 f}{\partial \mathcal{L}_m^2}.
\]

Introducing the energy-momentum tensor
\begin{equation}
    T_{\mu\nu}
    = g_{\mu\nu} \mathcal{L}_m
    - 2 \frac{\partial \mathcal{L}_m}{\partial g^{\mu\nu}} ,
\end{equation}
the metric field equations can be written as
\begin{align}\label{field}
      \frac{2}{\sqrt{-g}} \nabla_\alpha
      \left( f_Q \sqrt{-g} P^\alpha_{\;\;\mu\nu} \right)
      + f_Q \left(
      P_{\mu\alpha\beta} Q_\nu^{\;\;\alpha\beta}
      - 2 Q^{\alpha\beta}_{\;\;\;\mu} P_{\alpha\beta\nu}
      \right)
      \nonumber \\
      + \frac{1}{2} f g_{\mu\nu}
      =
      \frac{1}{2} f_{\mathcal{L}_m}
      \left( g_{\mu\nu} \mathcal{L}_m - T_{\mu\nu} \right).
\end{align}

If the variation is instead performed with respect to the affine connection, the corresponding field equation becomes
\begin{equation}
    \nabla_\mu \nabla_\nu
    \Bigl(
    4 \sqrt{-g}\, f_Q\, P^{\mu\nu}_{\;\;\;\;\alpha}
    + H_\alpha^{\;\;\mu\nu}
    \Bigr) = 0 .
    \label{conn}
\end{equation}

Here, $H_\alpha^{\;\;\mu\nu}$ denotes the hypermomentum tensor, which characterizes the coupling between matter and the affine connection. It is defined as
\begin{equation}
    H^{\alpha}{}_{\mu\nu}
    = \sqrt{-g}\, f_{\mathcal{L}_m}
    \frac{\delta \mathcal{L}_m}{\delta \Upsilon^{\alpha}{}_{\mu\nu}} .
\end{equation}

Now using \eqref{field} we can derive the energy balance equation like:
\begin{align}\label{balance}
    D_\mu T^\mu{}_\nu &= \frac{1}{f_{\mathcal{L}_m} \sqrt{-g}} \left[ \nabla_\alpha \nabla_\mu H^{\alpha \mu}{}_\nu - 2 Q_\mu \nabla_\alpha \left( \sqrt{-g} f_Q P^{\alpha \mu}{}_\nu \right) \right] \nonumber \\
    &\hspace{2.3cm}+ \frac{(\delta^{\mu}_{\nu}\mathcal{L}_m - T^{\mu}_{\nu})\partial_{\mu}(f_{\mathcal{L}_m})}{f_{\mathcal{L}_m}}.
\end{align}
    
The operator $D_\mu$ represents the covariant derivative associated with the Levi--Civita connection. Equation~\eqref{balance} explicitly indicates that the $f(Q,\mathcal{L}_m)$ theory leads to a non-conservation of the energy--momentum tensor. From this equation, one can straightforwardly derive the non-conserved continuity equation for the matter sector, which takes the form
\begin{equation}
    \dot{\rho} + 3H(\rho + p)
    = - \frac{\dot{f}_{\mathcal{L}_m}}{f_{\mathcal{L}_m}}
    \left( \mathcal{L}_m + \rho \right) .
\label{bcon}
\end{equation}

A detailed derivation of this equation~\eqref{balance} is provided in Appendix~B.

\subsection{Background Cosmological Dynamics}\label{sec: friedmann_derivation}

We consider the standard FLRW spacetime in our analysis,
\begin{equation}
    ds^2=-dt^2+a^2(t)(dx^2+dy^2+dz^2).
\end{equation}
 $a(t)$ generally denotes the cosmic scale factor and the expansion rate of the universe is followed by Hubble parameter, $H = \dot{a}/a$. We have considered the "perfect fluid" on which energy momentum tensor has the usual form $T_{\mu\nu}= (\rho+p)u_{\mu}u_{\nu} + pg_{\mu\nu}$. We can say about the matter density $\rho(=T^0_0)$ and pressure $p (= \frac{1}{3}T^i_i)$. Now implementing this steps in our  above field equation ~\eqref{field} , we can find the generalised Friedmann equation, which will look like:
\begin{eqnarray}
\label{eq:friedmann1}
    && 3H^2 =\frac{1}{4f_Q}\bigr[ f - f_{\mathcal{L}_m}(\rho + \mathcal{L}_m) \bigl],\\
     \label{eq:friedmannmain}
   && \dot{H} + 3H^2 + \frac{\dot{f_Q}}{f_Q}H=\frac{1}{4f_Q}\bigr[ f + f_{\mathcal{L}_m}(p - \mathcal{L}_m) \bigl]. 
\end{eqnarray}
These equations are more complex than the ones in standard Einstein gravity. They include terms that show how the "non-metricity" ($\mathcal{Q}$) and matter ($\mathcal{L}_m$) are mixed together. An important way to measure the expansion is through the "deceleration parameter" $\mathcal{Q}$. If $\mathcal{Q}$ is negative, the expansion is speeding up: 
\begin{equation}
    q = -1 - \frac{\dot{H}}{H^{2}}.
    \label{dec}
\end{equation}

To make these equations easier to read, we can group the new geometric terms into an effective density  ($\rho_{eff}$) and effective pressure ($p_{eff}$):
\begin{equation}\label{41}
3H^2=\rho_{eff}, \;\;\;\;\;  2\dot{H}+3H^2=-p_{eff}.
\end{equation}
This allows us to write a "conservation equation" which shows how energy shifts between matter and geometry as the Universe grows:
\begin{equation}\label{cons1}
\dot{\rho}_{eff}+3H\left(\rho_{eff}+p_{eff}\right)=0.
\end{equation}

\subsection{Proposed Cosmological Paradigm}

Building upon previous investigations, we propose the following generalized model:
\begin{equation}
    f(Q,L_m) = -Q^{\mu} + \alpha \mathcal{L}_m^{\nu} + \beta
    \label{choice}
\end{equation}

For the specific functional form given in Eq.~\eqref{choice}, the modified Friedmann equations take the following form:
\begin{align}
    3H^2 &= \bigg[\frac{\alpha \rho^\nu(2\nu-1)}{2^\mu(2\mu-1)}- \frac{\beta}{2^\mu(2\mu-1)}\bigg]^{\frac{1}{\mu}} \\
    2\mu\dot{H} + 3H^2 &= 3H^2 \bigg[\frac{\alpha \rho^\nu(\nu-1)-\beta}{(6H^2)^\mu(2\mu-1)}\bigg]
\end{align}

Substituting this into Eq.~\ref{eq:friedmann1} yields the following important relation: 
\begin{equation}\label{density}
    \alpha \rho^{\nu} = \frac{(6H^2)^\mu(2\mu-1) + \beta}{2\nu-1}
\end{equation}

Using this relation, we can derive the governing differential equation:
\begin{align}
    2\mu \frac{dH}{dt} + \frac{3\nu}{2\nu-1}H^2 = - \frac{\nu\beta (6H^2)^\mu}{2(2\nu-1)(2\mu-1)}
\end{align}

Furthermore, applying Eq.~\eqref{density} under specific parametric conditions allows us to express the Hubble parameter as a function of redshift: 
\begin{align}\label{model}
    H(z) = H_0 \left[\Gamma (1+z)^{\frac{3\nu}{2\nu-1}} + (1 -\Gamma)\right]^{\frac{1}{2\mu}}
\end{align}

Our parametrization is highlighted as 
\begin{equation}
\Gamma = \left(1 + \frac{\beta}{(6H_0^2)^\mu(2\mu-1)}\right).
\end{equation}
We have not treated $\mu$ as a free parameter; instead, we have proposed three models based on specific choices for $\mu$. 

\begin{itemize}
    \item \textbf{Model I:} We have chosen $\mu=1$, a case already studied in \cite{Myrzakulov:2025ycj}. However, in our context, it is treated with a different parametrization approach and an updated cosmographic analysis. 
    \item \textbf{Model II:} We have modified the geometry by adopting the choice $\mu = \nu$, which places the matter and geometry modifications on an equal power-law footing. 
    \item \textbf{Model III:} Our final model considers the choice $\mu = \frac{\nu}{2\nu - 1}$. This choice originates from the inherent mathematical symmetry of the Hubble function mentioned in Eq.~\eqref{model}.
\end{itemize}

As a gist, these selections allow us to explore the interplay between the non-metricity scalar and the matter Lagrangian under different geometric constraints.

\begin{equation}
\begin{aligned}
\text{Model I:} \quad   & f(\mathcal{Q},\mathcal{L}_m) = -\mathcal{Q} + \alpha \mathcal{L}_m^{\nu} + \beta \\[6pt]
\text{Model II:} \quad  & f(\mathcal{Q},\mathcal{L}_m) = -\mathcal{Q}^{\nu} + \alpha \mathcal{L}_m^{\nu} + \beta \\[6pt]
\text{Model III:} \quad & f(\mathcal{Q},\mathcal{L}_m) = -\mathcal{Q}^{\frac{\nu}{2\nu-1}} + \alpha \mathcal{L}_m^{\nu} + \beta
\end{aligned}
\end{equation}

To further investigate the cosmological implications of this model, we perform a cosmographic analysis, focusing on the evolution of the deceleration parameter, $q(z)$. The deceleration parameter is a crucial quantity that characterizes the expansion history of the universe, specifically indicating the transition between the decelerating and late-time accelerating phases. Using the relation derived in Eq.~\eqref{dec}, we can express the explicit redshift dependence of the deceleration parameter as follows:

\begin{equation}
    q(z) = -1 + \frac{ \left( \frac{3\nu}{2\mu(2\nu - 1)} \right) \Gamma (1+z)^{\frac{3\nu}{2\nu - 1}} }{ \Gamma (1+z)^{\frac{3\nu}{2\nu - 1}} + (1 - \Gamma) }
    \label{eq:qz_evolution}
\end{equation}

\begin{figure}[htbp]
    \centering
    \includegraphics[width=\linewidth]{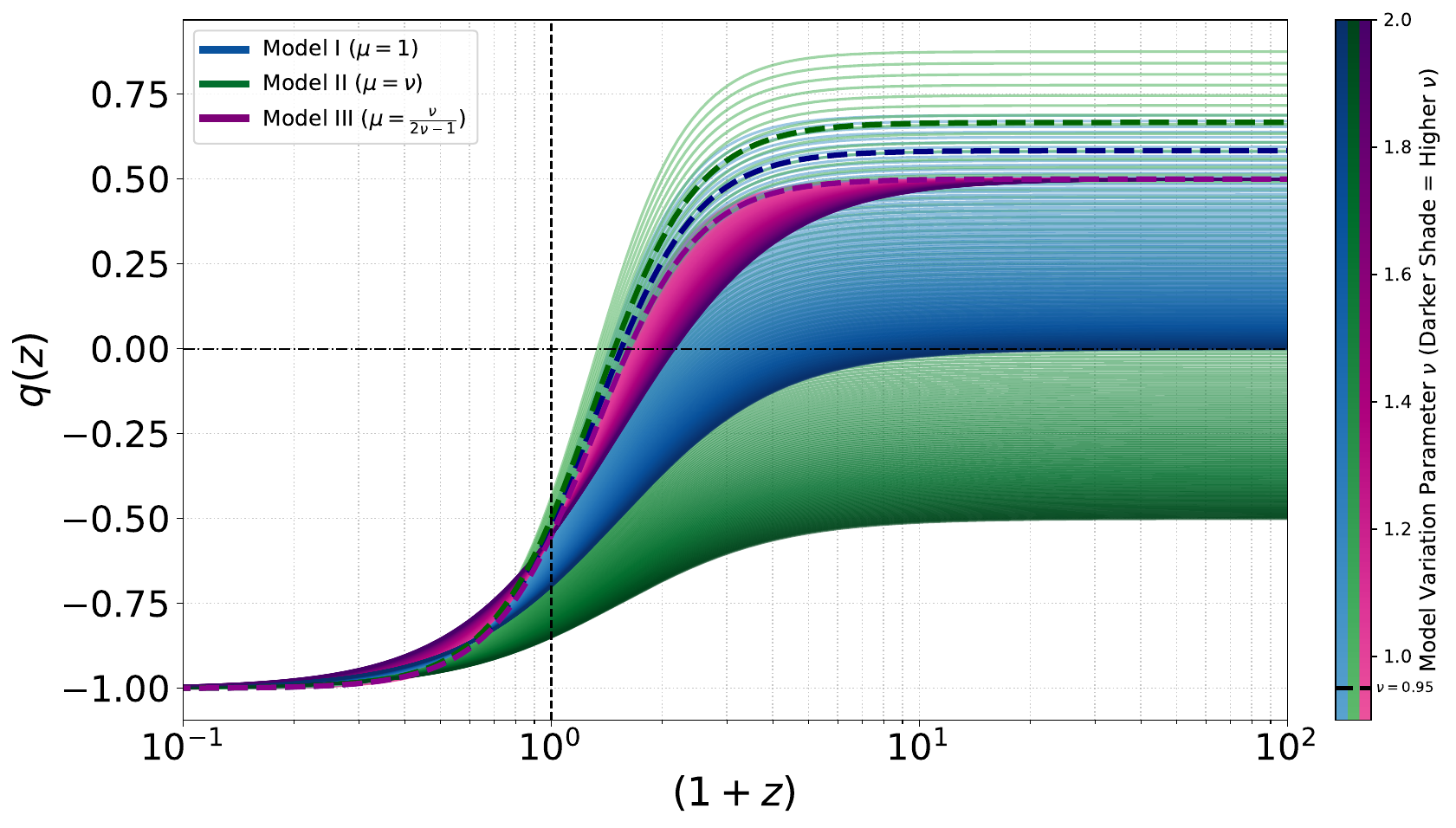}
    
    \vspace{0.5cm} 
    
    \includegraphics[width=\linewidth]{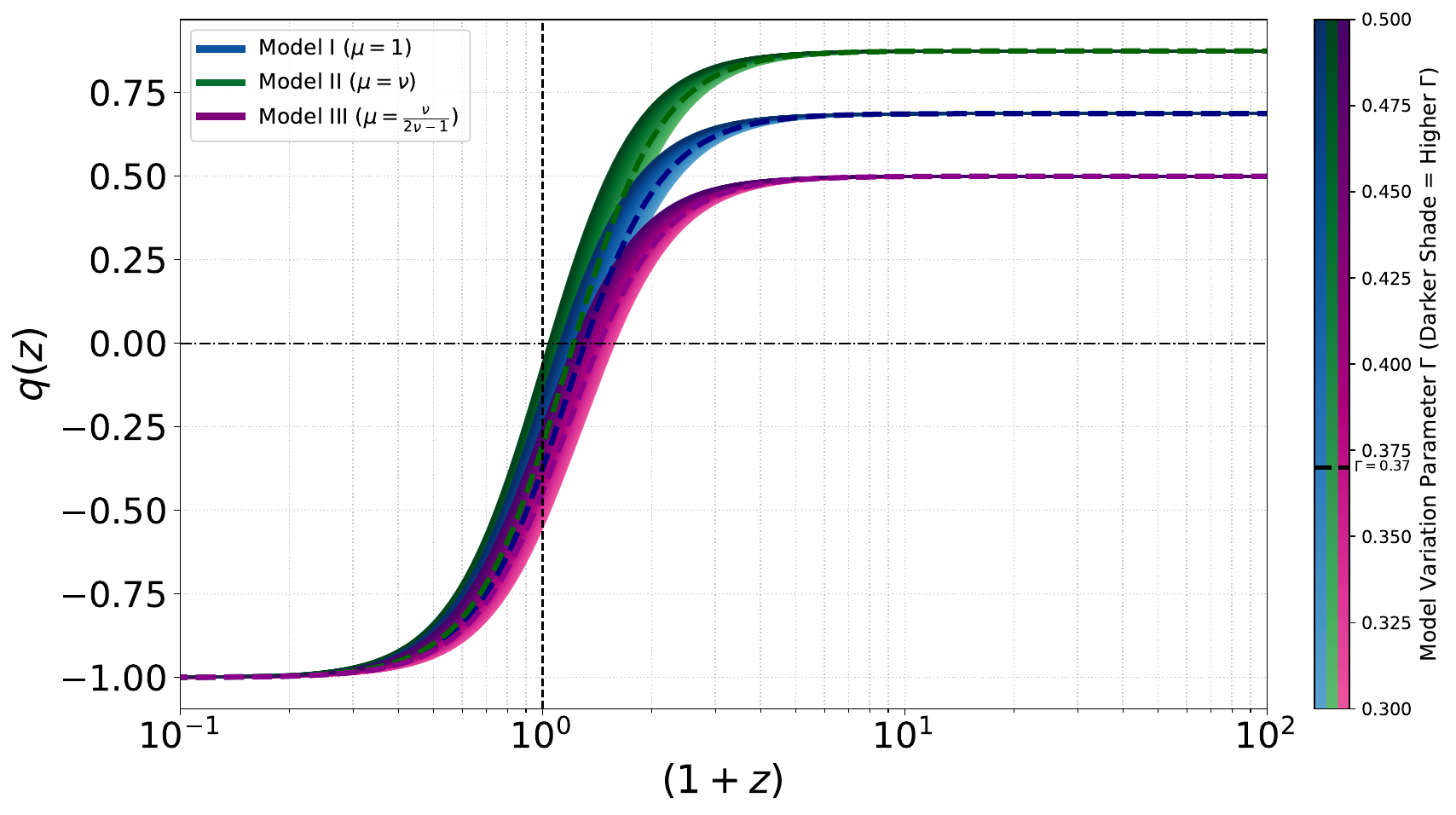}
    
    \caption{Cosmographic analysis of the proposed model. Top (a): Evolution of the deceleration parameter $q(z)$ for different values of the parameter $\nu$. Bottom (b): Evolution of $q(z)$ illustrating the effect of varying the parameter $\Gamma$.}
    \label{fig:cosmographic_analysis}
\end{figure}

The evolutionary behavior of the deceleration parameter is illustrated in Fig.~\ref{fig:cosmographic_analysis}.

In Figure~\ref{fig:cosmographic_analysis}, we study the evolution of the deceleration parameter, $q(z)$, across the three models. Because $q(z)$ depends on the cosmological parameters $\Gamma$ and $\nu$, we analyze its evolution with respect to variations in both. We observe that as $\nu$ increases, $q(z)$ enters the accelerating phase at a comparatively earlier epoch (higher $z$). Among the three models, the second model exhibits this transition the earliest. Conversely, the effect of $\Gamma$ is the opposite: larger values of $\Gamma$ delay the transition from the deceleration to the acceleration phase to a relatively later time (lower $z$). Once again, the second model is the most sensitive to these variations, although the overall effect induced by changing $\Gamma$ is relatively small.

\vspace{0.2cm}
\subsection {Scalar Perturbation and structure of Growth}

We execute scalar perturbation in this $f(\mathcal{Q},\mathcal{L_\text{m}})$ context.

\begin{align*}
    g_{\mu\nu} = \bar g_{\mu\nu} +\delta g_{\mu\nu} \\
    g^{\mu\nu} = \bar g^{\mu\nu} -\delta g^{\mu\nu}
\end{align*}

We have only considered the scalar degrees of freedom in the metric by introducing Bardeen potential $\psi$ and $\phi$ which reprents as, $\delta g_{00} = -2\psi$ and $\delta g_{ij} = -2\phi a^2\delta_{ij}$.

\begin{equation}
    dS^2= -(1+2\psi)dt^2 + (1 - 2\phi)a^2\delta_{ij}dx^idx^j
\end{equation}

We have considered Newtonian gauge in our treatment and we perturbed the matter sector as per usual prescription,
\begin{equation}
    \rho(x,t) = \bar\rho(t) + \delta \rho(x,t).
\end{equation}
Here $\bar\rho$ represents the background matter density and $\delta \rho$ is the perturbed part which we can expressed as $\delta \rho = \bar\rho \delta(x,t)$. As we have considered the perfect fluid so the components of the perturbed energy momentum tensor will be \cite{Joyce:2014kja},

\begin{align}
    \delta T^0_0 = -\delta\rho \\
    \delta T^0_i = (\bar\rho+\bar p)av_i\\
    \delta T^i_0 = -(\bar\rho+\bar p)\frac{v^i}{a}\\
    \delta T^i_j = \delta p \delta^{i}_{j}.
\end{align}

So now implementing the above conditions and perturbing the field equation ~\eqref{field} we gen get the perturbed Friedmann's equation, 

\begin{widetext}
\begin{align}
    \nabla^2\phi -6a^2H\bigg(\dot\phi +H\psi\bigg)\bigg(1 + \frac{12H^2f_{QQ}}{f_Q}\bigg) = -a^2\bigg [\bigg(\frac{f_{\mathcal{L}_m}}{4f_Q}\bigg)\delta \rho +\frac{f_{\mathcal{L}_m\mathcal{L}_m}}{4f_Q}\bigg(\bar{\mathcal{L}}_m + \bar\rho\bigg)\delta \mathcal{L}_m \bigg] \\
    \nabla^2\bigg(\dot\phi +H\psi\bigg)\bigg(1 + \frac{3H^2f_{QQ}}{f_Q}\bigg) = \bigg(\frac{f_{\mathcal{L}_m}}{4f_Q}\bigg) \bigg(\bar\rho + \bar p\bigg)a^2\theta - 3H\dot H \frac{f_{QQ}}{f_Q} \nabla^2\bigg(\psi + 3\phi\bigg)
\end{align}
\end{widetext}

In this case, we adopt the quasi-static approximation within sub-horizon scales in the absence of anisotropic stress ($\psi = \phi$). Since we have considered the matter Lagrangian to be $\mathcal{L}_m = \rho$ for the background dynamics, we also adopt $\delta \mathcal{L}_m = \delta \rho = \bar{\rho} \delta$ in the perturbative context. Consequently, we treat the perturbation of the matter Lagrangian as $\delta \mathcal{L}_m = \bar{\mathcal{L}}_m \delta$. By applying a Fourier transformation, we obtain the Poisson equation in $k$-space:,

\begin{align}
    k^2\psi = \frac{f_{\mathcal{L}_m}\bar\rho a^2}{4f_Q}\bigg(1 + \frac{\bar{\mathcal{L}}_m}{\bar\rho}(\bar{\mathcal{L}}_m + \bar\rho)\frac{f_{\mathcal{L}_m\mathcal{L}_m}}{f_{\mathcal{L}_m}}\bigg)\delta
    \label{poission}
\end{align}

To derive the evolution of matter perturbations, we first perturb the energy balance equation~\eqref{balance}. This yields the perturbed continuity and Euler equations, which are given by:
\begin{align}
    (2\nu-1)\dot\delta + \theta &= 0 \\
    \dot\theta + \left(2H - \frac{\dot{f}_{\mathcal{L}_m}}{f_{\mathcal{L}_m}}\right)\theta - \frac{k^2\psi}{a^2} + \frac{k^2\delta}{a^2}(\nu-1) &= 0
    \label{euc}
\end{align}
In this framework, the velocity divergence is defined as $\theta = \frac{1}{a}\partial^i v_i$. By combining equations~\eqref{poission} and \eqref{euc}, we can develop the second-order differential equation for the growth of matter perturbations:
\begin{equation}
    \ddot{\delta} + \left(2H - \frac{\dot{f}_{\mathcal{L}_m}}{f_{\mathcal{L}_m}} \right)\dot{\delta} + \left(\frac{f_{\mathcal{L}_m}\bar{\rho}}{4f_Q} - \frac{k^2}{a^2}\frac{(\nu-1)}{(2\nu-1)}\right)\delta = 0
\label{growth}
\end{equation}

Equation~\eqref{growth} describes the growth of structure in $f(Q,\mathcal{L}_m)$ gravity. The
effect of the matter sector $\mathcal{L}_m^{\nu}$ appears in both the coefficient of $\dot\delta$ and
the coefficient of $\delta$. To interpret these modifications, it is convenient to write the
equation in the generic form
\begin{equation}
\ddot\delta + A\,\dot\delta + B\,\delta = 0 ,
\end{equation}
where
\begin{equation}
A = 2H - \frac{\dot f_{\mathcal{L}_m}}{f_{\mathcal{L}_m}}, \qquad
B = \frac{f_{\mathcal{L}_m}\bar\rho}{4f_Q} - \frac{k^2}{a^2}\frac{\nu-1}{2\nu-1} .
\end{equation}
In $\Lambda$CDM these reduce to $A_{\Lambda{\rm CDM}}=2H$ and $B_{\Lambda{\rm CDM}}=-4\pi G\bar\rho$,
so that $\ddot\delta+2H\dot\delta-4\pi G\bar\rho\,\delta=0$. The coefficient $B$ acts as the source
term: for $\dot\delta=0$ one has $\ddot\delta=4\pi G\bar\rho\,\delta$, i.e.\ gravity accelerates the
collapse of overdense regions. The coefficient $A$ is the Hubble friction, through which the
expansion of the Universe damps the growth of perturbations.

In $f(Q,\mathcal{L}_m)$ gravity both coefficients are modified. The friction term becomes
$A = A_{\Lambda{\rm CDM}} - \dot f_{\mathcal{L}_m}/f_{\mathcal{L}_m}$, so that the friction is
reduced when $\dot f_{\mathcal{L}_m}/f_{\mathcal{L}_m}>0$ and enhanced when
$\dot f_{\mathcal{L}_m}/f_{\mathcal{L}_m}<0$. Using the background continuity equation, we find
\begin{equation}
\frac{\dot f_{\mathcal{L}_m}}{f_{\mathcal{L}_m}} = -\frac{3(\nu-1)}{2\nu-1}\,H ,
\end{equation}
so that, since $H>0$, the friction is enhanced for $\nu>1$ and reduced for $\nu<1$.

The coefficient $B$ contains two contributions. The first is the gravitational source term, the
analogue of $-4\pi G\bar\rho$, in which $f_Q$ and $f_{\mathcal{L}_m}$ define an effective
gravitational coupling. The second originates from the perturbation of $f_{\mathcal{L}_m}$,
$\partial^i\partial_j(\delta f_{\mathcal{L}_m}) = f_{\mathcal{L}_m\mathcal{L}_m}\bar\rho\,\nabla^2\delta$,
with $f_{\mathcal{L}_m\mathcal{L}_m}\equiv\partial f_{\mathcal{L}_m}/\partial\mathcal{L}_m$; it acts as a
scale-dependent effective pressure with sound speed $c^2_{\rm eff}=-(\nu-1)/(2\nu-1)$. For our action,
$f_Q=-\mu Q^{\mu-1}=-\mu(6H^2)^{\mu-1}<0$ and
$f_{\mathcal{L}_m}=\alpha\nu\mathcal{L}_m^{\nu-1}=\alpha\nu\bar\rho^{\nu-1}>0$ in all three models, so that
\begin{equation}
B = -\frac{\alpha\nu\,\bar\rho^{\nu}}{4\mu(6H^2)^{\mu-1}} - \frac{k^2}{a^2}\frac{\nu-1}{2\nu-1} .
\end{equation}
The first term is always negative and drives the growth. When $B<0$, perturbations grow; when
$B>0$, the effective pressure dominates and $\delta$ undergoes damped oscillations between
overdense ($\delta>0$) and underdense ($\delta<0$) states. Three cases can be distinguished:
\begin{itemize}
    \item \textbf{$\nu>1$:} both contributions to $B$ are negative, so $B<0$ and the perturbations
    grow, while the Hubble friction is enhanced. Since the scale-dependent term is amplified by
    $k^2/a^2H^2\sim10^5$ at $k=0.11\,h\,{\rm Mpc}^{-1}$, even a small departure $\nu-1$ leads to
    exponential growth, i.e.\ a gradient instability.

    \item \textbf{$\nu=1$:} all modifications related to $f_{\mathcal{L}_m}$ vanish, $\mu=1$ in all three
    models, and the growth equation reduces exactly to that of $\Lambda$CDM.

    \item \textbf{$1/2<\nu<1$:} the effective pressure is positive and the Hubble friction is
    reduced. The source term changes sign, $B>0$, when
    \begin{equation}
    \frac{k^2}{a^2}\frac{1-\nu}{2\nu-1} > \frac{\alpha\nu\,\bar\rho^{\nu}}{4\mu(6H^2)^{\mu-1}} ,
    \end{equation}
    in which case the perturbations oscillate instead of growing. Since $k^2/a^2H^2$ increases with
    time, this condition is first met at late times.
\end{itemize}

From the above discussion, we can define an effective Newtonian gravitational constant in
$f(Q,\mathcal{L}_m)$ gravity. The source term of Eq.~\eqref{growth} is modified through $f_Q$ and
$f_{\mathcal{L}_m}$; identifying it with its $\Lambda$CDM counterpart,
\begin{equation}
\frac{f_{\mathcal{L}_m}\bar\rho}{4f_Q} \equiv -4\pi G_{\rm eff}\,\bar\rho ,
\end{equation}
and recalling that $16\pi G=1$ in our action, we obtain
\begin{equation}
\frac{G_{\rm eff}}{G} = -\frac{f_{\mathcal{L}_m}}{f_Q}
= \frac{\alpha\nu\,\bar\rho^{\,\nu-1}}{\mu\,(6H^2)^{\mu-1}} .
\label{Geff}
\end{equation}
Since $f_Q<0$ and $f_{\mathcal{L}_m}>0$, $G_{\rm eff}/G$ is positive in all three models, so that gravity
remains attractive. In general, $G_{\rm eff}$ evolves with time through $\bar\rho$ and $H$, whereas for
$\mu=\nu=1$ it becomes constant and the growth equation reduces to the $\Lambda$CDM form. Note that the amplitude of $G_{\rm eff}$ depends on the constant $\alpha$, which is not fixed by the
cosmological data, since only the combination $\alpha\rho^{\nu}$ enters the background evolution
and is absorbed into $\Gamma$ in our parametrization.

To facilitate a direct comparison with cosmological observations, we transform the evolution equation~\eqref{growth} by changing the independent variable from cosmic time $t$ to the scale factor $a$. By utilizing the differential operator relation $d/dt = aH(a)d/da$, we can recast the growth of matter perturbations in terms of $\delta(a)$. The resulting second-order differential equation for the density contrast is given by:

\begin{align}
    \frac{d^2\delta}{da^2} + \frac{\mathcal{P}(a)}{a}\frac{d\delta}{da}
    &- \frac{1}{a^2}\Bigg[ \frac{3\nu(2\mu-1)\Gamma a^{-p}}{2\mu(2\nu-1)U} \nonumber \\
    &+ \frac{k^2(\nu-1)}{a^2H^2(2\nu-1)} \Bigg]\delta = 0
\end{align}

where we have defined:
\begin{align}
    \mathcal{P}(a) &= \frac{3 \left( \Gamma a^{-p} [2\mu(3\nu-2) - \nu] + 2\mu(3\nu-2)(1-\Gamma) \right)}{2\mu(2\nu-1)U} \\
    U(a) &= \Gamma a^{-p} + 1 - \Gamma, \quad p = \frac{3\nu}{2\nu-1}, \quad H^2(a) = H_0^2 U^{1/\mu}
\end{align}

Using the previous equation, we can calculate the linear growth rate, $f(a)$, defined as:
\begin{equation}
    f(a) = \frac{d \ln \delta}{d \ln a}
\end{equation}

To connect this theoretical rate to actual observations, we introduce the matter density fluctuation, defined as $\sigma_8(a) = \sigma_{8,0}\frac{\delta(a)}{\delta(a=1)}$. By combining these, we create the observable parameter $f\sigma_8(a)$. The resulting definition is:
\begin{equation}
    f\sigma_8(a) = \sigma_{8,0}\frac{a}{\delta(a=1)}\frac{d \delta}{da}
\end{equation}.

In our analysis, we adopt the Planck 2018 value $\sigma_{8,0} = 0.8111 \pm 0.0060$ (68\% CL) for the present-day amplitude of matter fluctuations \cite{Planck:2018vyg}. We can visualize the picture from the following plot related to the structure of growth, provided in section~\ref{sec-Result}.

\section{Results and Discussions}
\label{sec-Result}
In this section, we discuss the parameter constraints obtained for the three models, first from the
background expansion alone and then with the inclusion of the growth of structure. By confronting
our models with the observational datasets, we obtain constraints on the model parameters $H_0$,
$\Gamma$ and $\nu$. These results allow us to assess how well the $f(Q,\mathcal{L}_m)$ gravity
framework describes both the expansion history of the Universe and the growth of structure. The
analysis is performed first with the background combinations OHD+DESI+SNe Ia, and then with the
growth-of-structure data added, i.e.\ OHD+DESI+SNe Ia+$f\sigma_8$, where SNe Ia denotes one of the three
samples Union3, Pantheon+ (PP) and DESY5 (Dovekie). The resulting constraints at 68\% and 95\% CL are
listed in Table~\ref{tab:constraints}.

The Hubble constant $H_0$ is a key parameter, as it sets the present expansion rate of the Universe.
For the background combinations OHD+DESI+Union3, OHD+DESI+PP and OHD+DESI+DESY5, we find
$H_0\simeq67.4$--$67.9\,{\rm km\,s^{-1}Mpc^{-1}}$ with uncertainties of $\simeq0.5$--$0.7$ in all three
models, consistent with the Planck 2018 $\Lambda$CDM value, $H_0=67.36\pm0.54\,{\rm km\,s^{-1}Mpc^{-1}}$
\cite{Planck:2018vyg}.

For all background combinations OHD+DESI+SNe Ia, we find $\Gamma$ in the range $0.356$--$0.373$.
We now turn to the parameter $\nu$. For the background combinations OHD+DESI+SNe Ia, Models I and II
constrain it to the range $1.029$--$1.061$, and exclude $\nu=1$ at 95\% CL. In Model III, the values are
considerably larger, in the range $1.309$--$1.42$, and again exclude $\nu=1$ at 95\% CL.

We now discuss the effect of including the $f\sigma_8$ data, which changes the picture
considerably. For each SNe Ia sample, the three models yield practically identical constraints on
$H_0$, $\Gamma$ and $\nu$ (Table~\ref{tab:constraints}). We find
$H_0=69.06\pm0.43$, $68.93\pm0.41$ and $68.82\pm0.40\,{\rm km\,s^{-1}Mpc^{-1}}$ for Union3, PP and DESY5,
respectively. The parameter $\Gamma$ lies in the ranges $0.2941$--$0.2942$ (Union3),
$0.2963$--$0.2964$ (PP) and $0.2983$--$0.2984$ (DESY5), with nearly identical uncertainties at both
68\% and 95\% CL. Most notably, $\nu$ is confined to unity, $\nu=0.99999957$--$0.99999965$ with
$\sigma_\nu\simeq5.2\times10^{-7}$, so that $\nu=1$ lies within $1\sigma$ for all three SNe samples.

The physical origin of this tight constraint can be understood from the growth equation,
Eq.~\eqref{growth}. For $\nu\neq1$, the non-linear matter sector $\alpha\mathcal{L}_m^{\nu}$ makes
$f_{\mathcal{L}_m}$ depend on the local matter density, so that a density perturbation also perturbs
the coupling itself. This acts on the matter as an effective pressure with sound speed
$c^2_{\rm eff}=-(\nu-1)/(2\nu-1)$, which, unlike the gravitational source term, grows with the wavenumber
as $k^2/a^2H^2$ and is therefore strongly enhanced on the sub-horizon scales probed by RSD. For
$\nu>1$ the effective pressure is negative: instead of resisting the collapse of overdense regions,
it adds to gravity, and the perturbations grow in a runaway manner. For $\nu<1$ the pressure is
positive and opposes gravity, suppressing the growth and eventually turning it into oscillations
between overdense and underdense states. The observed growth rate, however, follows closely the
$\Lambda$CDM behaviour of gravitationally driven, monotonic growth. The RSD data therefore tolerate
only a negligible effective pressure, which forces $\nu$ to unity with a precision of $\sim10^{-6}$.

\begin{figure*}
    \centering
    \includegraphics[width= 1.0\linewidth]{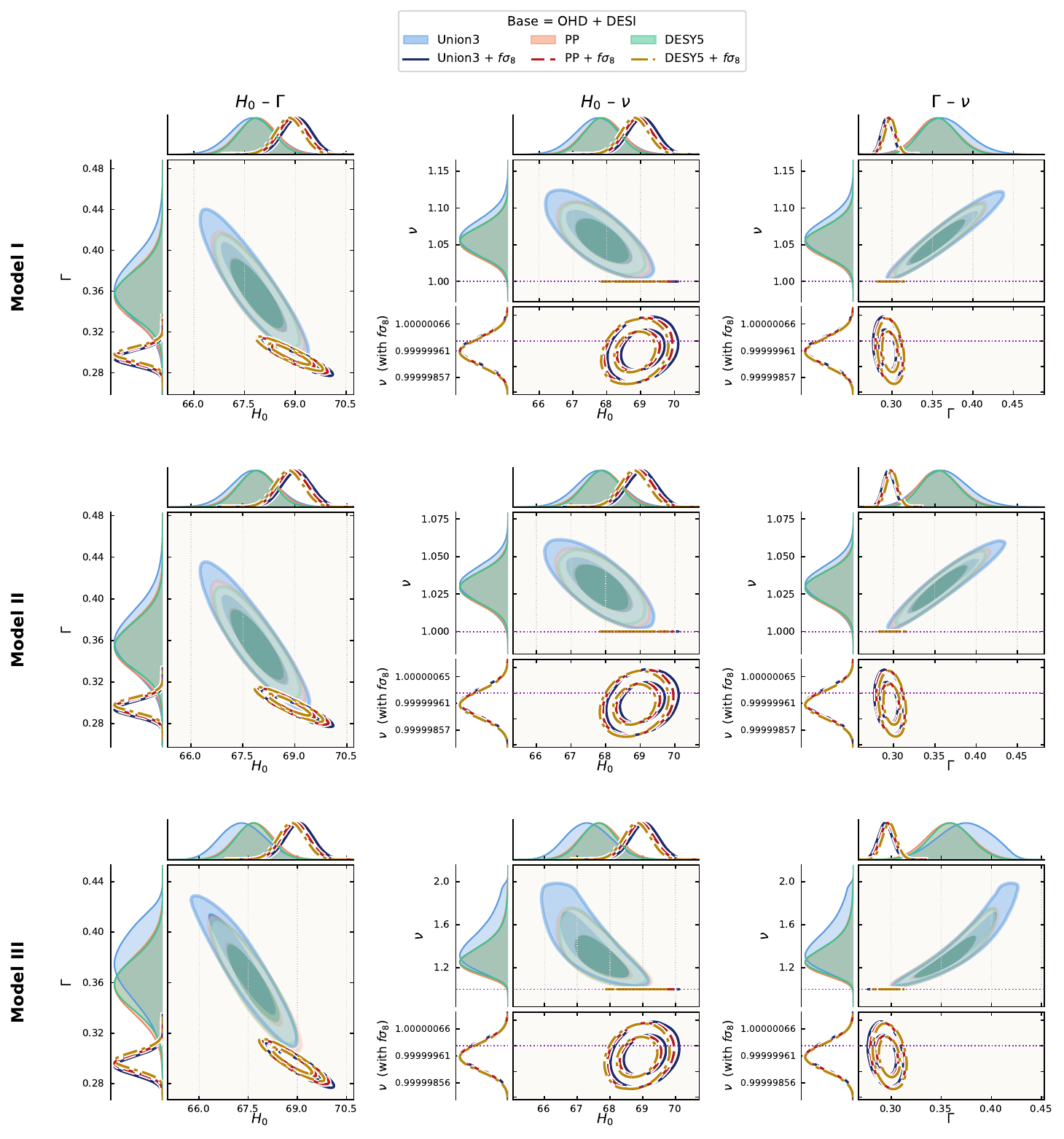}
    \caption{2D contour plots with 1D posterior distributions for different combinations of OHD, DESI, and supernova datasets.}
    \label{2Dcontour}
\end{figure*}

Since $\mu\to1$ as $\nu\to1$ in all three models, their background and perturbation equations become
indistinguishable, which explains the identical constraints. In this limit the parameter $\Gamma$
acquires a direct physical meaning: for $\mu=\nu=1$ the constant $\beta$ plays the role of a
cosmological constant, $\beta=-2\Lambda$, and the definition of $\Gamma$ gives
\begin{equation}
\Gamma = 1 + \frac{\beta}{6H_0^2} = 1 - \frac{\Lambda}{3H_0^2} = 1 - \Omega_{\Lambda} = \Omega_{m0} ,
\end{equation}
for a spatially flat Universe, while the Hubble rate reduces to the $\Lambda$CDM form
$H^2=H_0^2\left(\Gamma a^{-3}+1-\Gamma\right)$. Thus $\Gamma$ coincides with the present matter density
parameter, and the values $\Gamma\simeq0.294$--$0.298$ are close to the Planck 2018 estimate
$\Omega_{m0}=0.315$ \cite{Planck:2018vyg}. The higher values $\Gamma\simeq0.36$ obtained from the
background data alone reflect the positive correlation between $\Gamma$ and $\nu$
(Fig.~\ref{2Dcontour}): once $\nu$ is pinned to unity, $\Gamma$ decreases and, through its
anti-correlation with $H_0$, $H_0$ increases by $\simeq1\,{\rm km\,s^{-1}Mpc^{-1}}$.

\begin{figure*}[htbp]
    \centering
    \includegraphics[width=\textwidth]{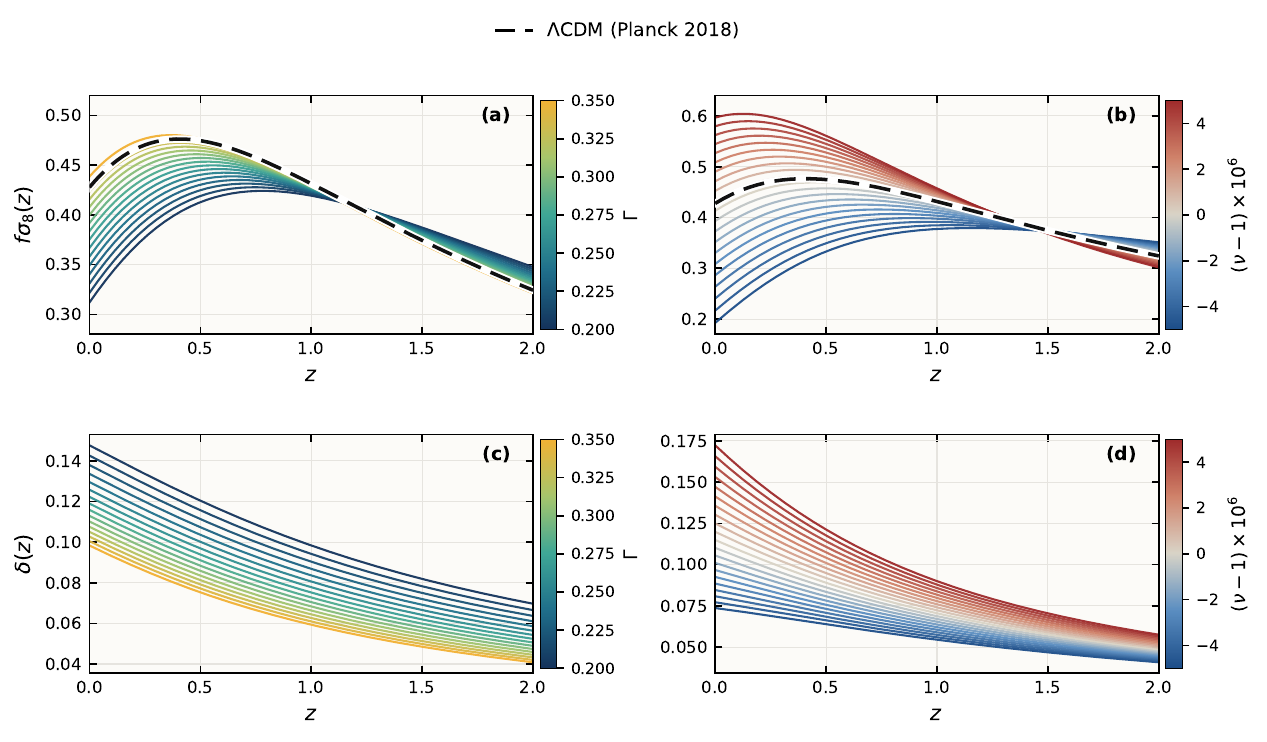}
    \caption{Growth of matter perturbations around the constraints of Table~\ref{tab:constraints},
    valid for Models I, II and III, whose predictions coincide to better than $10^{-6}$.
    Top: $f\sigma_8(z)$ when varying (a) $\Gamma\in[0.20,0.35]$ and (b) $\nu$ around unity, compared
    with $\Lambda$CDM (Planck 2018, $\Omega_{m0}=0.3153$, dashed). Bottom: the corresponding density
    contrast $\delta(z)$ for (c) $\Gamma$ and (d) $\nu$, integrated from $a_i=10^{-3}$ with
    $\phi_i=-6\times10^{-7}$; its amplitude is set by this normalisation and cancels in $f\sigma_8$.}
    \label{fig:growth}
\end{figure*}

We now visualise the growth of structure through the evolution of $f\sigma_8(z)$, shown in
Fig.~\ref{fig:growth}. Since $\nu$ is driven to unity in the perturbative regime, all three models
behave identically, and a single set of curves applies to Models I, II and III. A higher value of
$\Gamma$ increases $f\sigma_8$ at low redshift, while around $z\sim1.2$ this behaviour is reversed.
In contrast, $f\sigma_8$ is extremely sensitive to $\nu$: even a departure of
$|\nu-1|\sim5\times10^{-6}$ shifts $f\sigma_8(z=0)$ from $\simeq0.2$ to $\simeq0.6$. The density
contrast shows the same trend (panels c and d): it grows monotonically in all cases, while the
effect of $\nu$ appears only at late times, when the scale-dependent effective pressure becomes
important. This explains why the RSD data restrict $\nu$ to $|\nu-1|\lesssim10^{-6}$, within which
the growth remains close to the $\Lambda$CDM prediction.

\begin{table*}
\centering
\resizebox{1.0\textwidth}{!}{%
\huge
\renewcommand{\arraystretch}{1.5}
\begin{tabular}{|c|c|c|c|c|c|c|}
\hline
\textbf{Parameter}
& ~\textbf{OHD+DESI+Union3}~
& ~\textbf{OHD+DESI+PP}~
& ~\textbf{OHD+DESI+DESY5}~
& ~\textbf{OHD+DESI+Union3+\boldmath{$f\sigma_8$}}~
& ~\textbf{OHD+DESI+PP+\boldmath{$f\sigma_8$}}~
& ~\textbf{OHD+DESI+DESY5+\boldmath{$f\sigma_8$}}~ \\
\hline\hline

\multicolumn{7}{|c|}{\textbf{Model I}} \\
\hline
$H_0$ & $67.74^{+0.65(+1.27)}_{-0.65(-1.28)}$ & $67.87^{+0.56(+1.10)}_{-0.56(-1.09)}$ & $67.84^{+0.51(+1.01)}_{-0.51(-1.01)}$ & $69.06^{+0.43(+0.84)}_{-0.42(-0.84)}$ & $68.93^{+0.41(+0.82)}_{-0.41(-0.82)}$ & $68.82^{+0.40(+0.80)}_{-0.40(-0.79)}$ \\

$\Gamma_{\rm}$ & $0.365^{+0.027(+0.061)}_{-0.032(-0.055)}$ & $0.357^{+0.023(+0.049)}_{-0.025(-0.046)}$ & $0.358^{+0.021(+0.046)}_{-0.023(-0.042)}$ & $0.2942^{+0.0072(+0.0145)}_{-0.0072(-0.0138)}$ & $0.2964^{+0.0070(+0.0139)}_{-0.0070(-0.0135)}$ & $0.2983^{+0.0068(+0.0136)}_{-0.0068(-0.0132)}$ \\

$\nu$ & $1.061^{+0.023(+0.051)}_{-0.026(-0.046)}$ & $1.055^{+0.019(+0.043)}_{-0.022(-0.039)}$ & $1.056^{+0.018(+0.041)}_{-0.021(-0.038)}$ & $0.99999965^{+0.00000052(+0.00000104)}_{-0.00000052(-0.00000101)}$ & $0.99999961^{+0.00000052(+0.00000103)}_{-0.00000052(-0.00000101)}$ & $0.99999957^{+0.00000052(+0.00000103)}_{-0.00000052(-0.00000101)}$ \\

\hline\hline

\multicolumn{7}{|c|}{\textbf{Model II}} \\
\hline
$H_0$ & $67.82^{+0.64(+1.26)}_{-0.64(-1.26)}$ & $67.91^{+0.55(+1.09)}_{-0.55(-1.09)}$ & $67.89^{+0.51(+1.00)}_{-0.51(-0.99)}$ & $69.06^{+0.43(+0.84)}_{-0.42(-0.83)}$ & $68.93^{+0.42(+0.82)}_{-0.42(-0.82)}$ & $68.82^{+0.41(+0.79)}_{-0.40(-0.79)}$ \\

$\Gamma_{\rm }$ & $0.361^{+0.027(+0.060)}_{-0.031(-0.054)}$ & $0.356^{+0.023(+0.049)}_{-0.025(-0.045)}$ & $0.356^{+0.021(+0.045)}_{-0.023(-0.042)}$ & $0.2942^{+0.0072(+0.0143)}_{-0.0071(-0.0138)}$ & $0.2963^{+0.0070(+0.0140)}_{-0.0070(-0.0136)}$ & $0.2983^{+0.0068(+0.0137)}_{-0.0068(-0.0132)}$ \\

$\nu$ & $1.032^{+0.012(+0.024)}_{-0.012(-0.024)}$ & $1.029^{+0.011(+0.021)}_{-0.011(-0.021)}$ & $1.030^{+0.010(+0.020)}_{-0.010(-0.020)}$ & $0.99999965^{+0.00000052(+0.00000103)}_{-0.00000052(-0.00000101)}$ & $0.99999961^{+0.00000052(+0.00000104)}_{-0.00000052(-0.00000101)}$ & $0.99999957^{+0.00000052(+0.00000104)}_{-0.00000052(-0.00000100)}$ \\

\hline\hline

\multicolumn{7}{|c|}{\textbf{Model III}} \\
\hline
$H_0$ & $67.36^{+0.62(+1.35)}_{-0.71(-1.23)}$ & $67.68^{+0.57(+1.14)}_{-0.58(-1.12)}$ & $67.66^{+0.53(+1.04)}_{-0.53(-1.03)}$ & $69.06^{+0.43(+0.84)}_{-0.42(-0.84)}$ & $68.93^{+0.42(+0.82)}_{-0.42(-0.82)}$ & $68.82^{+0.41(+0.80)}_{-0.41(-0.79)}$ \\

$\Gamma_{\rm }$ & $0.373^{+0.029(+0.048)}_{-0.024(-0.049)}$ & $0.358^{+0.022(+0.043)}_{-0.022(-0.043)}$ & $0.359^{+0.020(+0.040)}_{-0.020(-0.039)}$ & $0.2941^{+0.0072(+0.0144)}_{-0.0072(-0.0139)}$ & $0.2964^{+0.0070(+0.0141)}_{-0.0070(-0.0135)}$ & $0.2984^{+0.0069(+0.0137)}_{-0.0069(-0.0132)}$ \\

$\nu$ & $1.42^{+0.14(+0.44)}_{-0.26(-0.35)}$ & $1.309^{+0.094(+0.318)}_{-0.181(-0.263)}$ & $1.314^{+0.092(+0.302)}_{-0.171(-0.253)}$ & $0.99999965^{+0.00000052(+0.00000104)}_{-0.00000052(-0.00000101)}$ & $0.99999961^{+0.00000052(+0.00000104)}_{-0.00000052(-0.00000100)}$ & $0.99999957^{+0.00000052(+0.00000104)}_{-0.00000052(-0.00000101)}$ \\

\hline

\end{tabular} }
\caption{Constraints on the model parameters at 68\% (95\%) CL. The $f\sigma_8$ columns use RSD data at $k=0.11\,h\,{\rm Mpc}^{-1}$.}
\label{tab:constraints}
\end{table*}

\section{Bayesian Evidence and Model Comparison}
The $\Lambda$CDM model provides a successful description of the Universe, although it still faces
several known limitations. Any alternative model must therefore be tested not only for its ability to
fit the data but also for its statistical economy, which can be quantified through the Bayesian
evidence.

For a model $M$ with parameters $\theta$ and data $x$, Bayes' theorem reads
\begin{equation}
p(\theta|x, M) = \frac{p(x|\theta, M)\,\pi(\theta|M)}{p(x|M)},
\label{eq:bayes}
\end{equation}
where $p(x|\theta, M)$ is the likelihood, $\pi(\theta|M)$ is the prior, and the normalisation
\begin{equation}
Z \equiv p(x|M) = \int p(x|\theta, M)\,\pi(\theta|M)\,d\theta
\end{equation}
is the Bayesian evidence, i.e.\ the likelihood averaged over the prior. To compare two models $M_i$
and $M_j$ we use the Bayes factor
\begin{equation}
B_{ij} = \frac{p(x|M_i)}{p(x|M_j)} = \frac{Z_i}{Z_j},
\qquad
\ln B_{ij} \equiv \Delta\ln Z = \ln Z_i - \ln Z_j .
\label{eq:bayes_factor}
\end{equation}
Throughout this work $M_i$ is $\Lambda$CDM and $M_j$ is the model under consideration, so that
$\Delta\ln Z = \ln Z_{\Lambda{\rm CDM}} - \ln Z_{\rm Model}$, and a negative value of $\Delta\ln Z$
indicates that the data favour the extended model. We interpret $|\Delta\ln Z|$ on the Jeffreys
scale \cite{Trotta:2008qt}: the evidence is inconclusive for $|\Delta\ln Z|<1$, weak for
$1\le|\Delta\ln Z|<2.5$, moderate for $2.5\le|\Delta\ln Z|<5$, and strong for $|\Delta\ln Z|\ge5$.

\begin{figure*}[htbp]
    \centering
    \includegraphics[width=\textwidth]{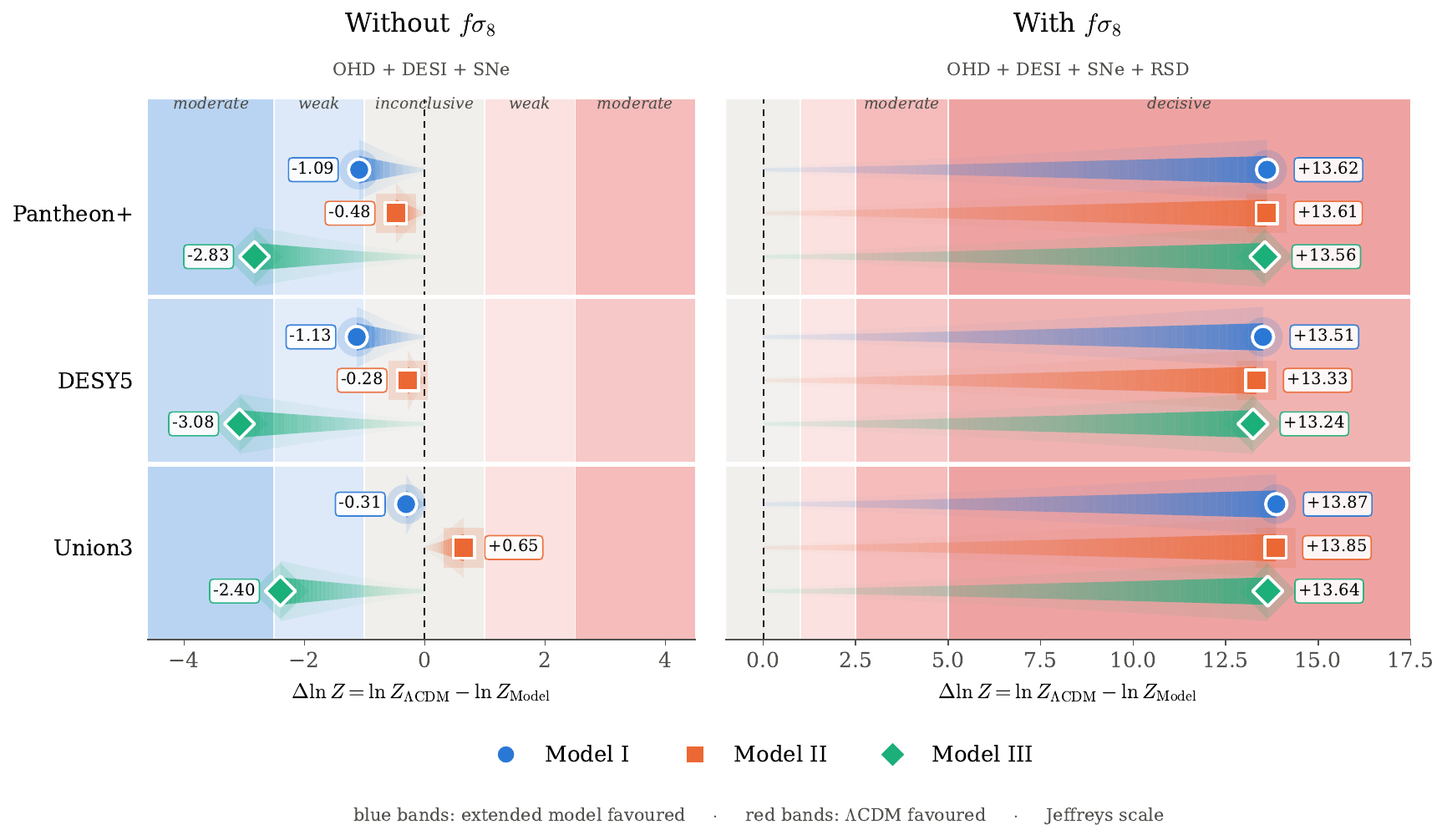}
    \caption{Bayesian model comparison on the Jeffreys scale for Models I, II and III relative to
    $\Lambda$CDM, without (left) and with (right) the $f\sigma_8$ data. The horizontal axis shows
    $\Delta\ln Z = \ln Z_{\Lambda{\rm CDM}} - \ln Z_{\rm Model}$: positive values (red bands) favour
    $\Lambda$CDM, while negative values (blue bands) favour the extended models.}
    \label{fig:bayesian_evidence}
\end{figure*}

The evidences were computed with the nested-sampling code \textbf{DYNESTY}\footnote{https://github.com/joshspeagle/dynesty.git}
\cite{Speagle:2019ivv}. The resulting values of $\Delta\ln Z$ for Models I, II and III, together
with the maximum log-likelihoods, are listed in Table~\ref{tab:BE}, and a graphical representation
on the Jeffreys scale is shown in Fig.~\ref{fig:bayesian_evidence}.

With background data alone, all three models improve the fit over $\Lambda$CDM, with
$\Delta\chi^2\simeq-6.5$ to $-9.1$. Model III receives moderate support from PP
($\Delta\ln Z=-2.83$) and DESY5 ($\Delta\ln Z=-3.08$) and weak support from Union3
($\Delta\ln Z=-2.40$); Model I is weakly favoured by PP and DESY5 and inconclusive for Union3, while
Model II remains inconclusive for all three SNe samples. Since the improvement in the fit is similar
for the three models, these differences originate mainly from the Occam factor: the evidence weighs
the gain in likelihood against the fraction of the prior volume of $\nu$ that the data actually
use \cite{Trotta:2008qt}. In Model III the posterior of $\nu$ is broad, so that a larger part of the
prior volume remains compatible with the data and the penalty is smaller, whereas in Model II $\nu$
is constrained most tightly, and the comparable gain in likelihood is largely offset by the penalty
for the extra parameter.

The inclusion of the $f\sigma_8$ data reverses this picture. All three models are now decisively
disfavoured with respect to $\Lambda$CDM, with nearly identical values
$\Delta\ln Z\simeq13.2$--$13.9$ for all SNe samples. This does not reflect a poorer description of the
data: the maximum likelihoods of the three models are practically identical to that of $\Lambda$CDM,
$|\Delta\chi^2|\lesssim0.7$. As discussed above, the $f\sigma_8$ data force $\nu$ to unity in order to
suppress the scale-dependent effective pressure; consequently $\mu\to1$ and $\Gamma\to\Omega_{m0}$ in
all three models, their distinctive features disappear, and each model reduces to $\Lambda$CDM. The
extra parameter $\nu$ then brings no improvement in the fit, while the data confine it to a width
$\sigma_\nu\simeq5.2\times10^{-7}$, about six orders of magnitude narrower than its prior range.
The Bayesian evidence penalises this unused prior volume, an effect known as the Occam penalty
\cite{Trotta:2008qt}. Since $\Lambda$CDM is nested in each model at $\nu=1$, the Laplace
approximation (see, e.g., Eq.~(26) of Ref.~\cite{Trotta:2008qt}) for a flat prior of width
$\Delta\nu_{\rm prior}$ gives
\begin{equation}
\Delta\ln Z \simeq -\Delta\ln\mathcal{L}_{\max}
+ \ln\!\left(\frac{\Delta\nu_{\rm prior}}{\sqrt{2\pi}\,\sigma_\nu}\right),
\label{eq:occam}
\end{equation}
where $\Delta\ln\mathcal{L}_{\max}\equiv\ln\mathcal{L}_{\max}^{\rm Model}-\ln\mathcal{L}_{\max}^{\Lambda{\rm CDM}}$.
The first term rewards the improvement in the fit, while the second is the Occam penalty.

where the first term rewards the improvement in the fit and the second is the Occam penalty. With the
flat prior $\nu\in[0.6,2]$ ($\Delta\nu_{\rm prior}=1.4$), the Occam term is $\simeq13.9$ in all cases.
Together with $\Delta\ln\mathcal{L}_{\max}\simeq0.3$, this reproduces the nested-sampling results, and
the Savage--Dickey density ratio \cite{Trotta:2005ar} evaluated from the chains gives consistent
values, $\ln B\simeq13.5$--$13.7$. Hence the $f\sigma_8$ data do not disfavour the background
expansion of these models; rather, they show that the additional freedom introduced by $\nu$ is not
required once the growth of structure is taken into account.

\begin{table*}[htbp]
\centering
\renewcommand{\arraystretch}{1.3}
\setlength{\tabcolsep}{8pt}
\begin{tabular}{l l c c c c c}
\toprule[1.2pt]
\textbf{Dataset Combination} & \textbf{Model}
& $\boldsymbol{\ln\mathcal{L}_{\max}^{\text{Model}}}$ & $\boldsymbol{\ln\mathcal{L}_{\max}^{\Lambda\text{CDM}}}$
& $\boldsymbol{\ln Z_{\text{Model}}}$ & $\boldsymbol{\ln Z_{\Lambda\text{CDM}}}$ & $\boldsymbol{\Delta \ln Z}$ \\
\midrule[0.8pt]
\rowcolor{gray!15}
\multicolumn{7}{c}{\textbf{Combinations Without $f\sigma_8$}} \\
\midrule
\multirow{3}{*}{OHD+DESI+Pantheon+}
 & Model I   & $-726.67$ & \multirow{3}{*}{$-730.34$} & $-743.15$ & \multirow{3}{*}{$-744.24$} & $-1.09$ \\
 & Model II  & $-726.72$ & & $-743.76$ & & $-0.48$ \\
 & Model III & $-726.54$ & & $-741.41$ & & $-2.83$ \\
\midrule
\multirow{3}{*}{OHD+DESI+DESY5}
 & Model I   & $-840.91$ & \multirow{3}{*}{$-845.21$} & $-852.93$ & \multirow{3}{*}{$-854.07$} & $-1.13$ \\
 & Model II  & $-840.98$ & & $-853.79$ & & $-0.28$ \\
 & Model III & $-840.63$ & & $-850.98$ & & $-3.08$ \\
\midrule
\multirow{3}{*}{OHD+DESI+Union3}
 & Model I   & $-39.01$ & \multirow{3}{*}{$-42.32$} & $-50.64$ & \multirow{3}{*}{$-50.95$} & $-0.31$ \\
 & Model II  & $-39.09$ & & $-51.60$ & & $0.65$ \\
 & Model III & $-38.61$ & & $-48.55$ & & $-2.40$ \\
\midrule[0.8pt]
\rowcolor{gray!15}
\multicolumn{7}{c}{\textbf{Combinations Including $f\sigma_8$}} \\
\midrule
\multirow{3}{*}{OHD+DESI+$f\sigma_8$+Pantheon+}
 & Model I   & $-737.44$ & \multirow{3}{*}{$-737.74$} & $-764.74$ & \multirow{3}{*}{$-751.12$} & $13.62$ \\
 & Model II  & $-737.44$ & & $-764.73$ & & $13.61$ \\
 & Model III & $-737.44$ & & $-764.68$ & & $13.56$ \\
\midrule
\multirow{3}{*}{OHD+DESI+$f\sigma_8$+DESY5}
 & Model I   & $-852.30$ & \multirow{3}{*}{$-852.65$} & $-875.03$ & \multirow{3}{*}{$-861.52$} & $13.51$ \\
 & Model II  & $-852.30$ & & $-874.85$ & & $13.33$ \\
 & Model III & $-852.30$ & & $-874.76$ & & $13.24$ \\
\midrule
\multirow{3}{*}{OHD+DESI+$f\sigma_8$+Union3}
 & Model I   & $-49.42$ & \multirow{3}{*}{$-49.66$} & $-72.32$ & \multirow{3}{*}{$-58.45$} & $13.87$ \\
 & Model II  & $-49.42$ & & $-72.30$ & & $13.85$ \\
 & Model III & $-49.42$ & & $-72.09$ & & $13.64$ \\
\bottomrule[1.2pt]
\end{tabular}
\caption{Maximum log-likelihood ($\ln\mathcal{L}_{\max}$) and Bayesian evidence ($\ln Z$) of Models I, II and III
and of $\Lambda$CDM for the six dataset combinations, with
$\Delta\ln Z = \ln Z_{\Lambda{\rm CDM}} - \ln Z_{\rm Model}$; positive values favour $\Lambda$CDM.}
\label{tab:BE}
\end{table*}

\section{Summary and Conclusions}
\label{sec-summary}

In this work we have investigated a generalised additive form of $f(Q,\mathcal{L}_m)$ gravity, in which the non-metricity sector and the matter Lagrangian enter the action without cross terms, so that geometry and matter are minimally coupled. Even in this case, a non-linear dependence on $\mathcal{L}_m$ ($\nu\neq1$) renders $f_{\mathcal{L}_m}$ time dependent, so that the matter sector is no longer separately conserved and both the expansion history and the growth of structure depart from their $\Lambda$CDM counterparts. We considered three realisations of the model, Model I ($\mu=1$), Model II ($\mu=\nu$) and Model III ($\mu=\nu/(2\nu-1)$), all of which contain $\Lambda$CDM as the nested limit $\nu=1$. We derived the background evolution and the equation governing the linear matter density contrast, and confronted the models with cosmic chronometers, DESI DR2 BAO and three Type Ia supernova compilations, Pantheon+, Union3 and DESY5 (Dovekie recalibration), both with and without redshift-space distortion measurements of $f\sigma_8$. Parameter constraints were obtained through MCMC analyses, and the statistical preference relative to $\Lambda$CDM was quantified through the Bayesian evidence computed with nested sampling. Our main findings can be summarised as follows.

\begin{itemize}
    \item \textbf{Background level.} With background data alone, all three models provide a better fit than $\Lambda$CDM, with $\Delta\chi^2\simeq-6.5$ to $-9.1$, and the data prefer a departure from the linear matter Lagrangian, $\nu>1$; for instance, with DESY5 we find $\nu\simeq1.06$, $1.03$ and $1.31$ for Models I, II and III, respectively. The corresponding values of $H_0$ remain close to the Planck-like range, $H_0\simeq67.7$--$67.9~{\rm km\,s^{-1}\,Mpc^{-1}}$, while $\Gamma\simeq0.36$.

    \item \textbf{Bayesian evidence from the background.} Model III is moderately favoured over $\Lambda$CDM by Pantheon+ ($\Delta\ln Z=-2.83$) and DESY5 ($\Delta\ln Z=-3.08$), and weakly by Union3 ($\Delta\ln Z=-2.40$). Model I is weakly favoured by Pantheon+ and DESY5, whereas Model II remains inconclusive for all SNe samples. Since the improvement in the fit is comparable for the three models, these differences arise mainly from the Occam factor, i.e.\ from how much of the prior volume of $\nu$ remains compatible with the data.

    \item \textbf{Growth of structure.} At the perturbative level the non-linear matter coupling generates a scale-dependent effective pressure, $c^2_{\rm eff}=-(\nu-1)/(2\nu-1)$, together with a modified friction term in the growth equation. Since its contribution is enhanced by $(k/aH)^2$ on sub-horizon scales, even a small departure from $\nu=1$ leads either to a runaway growth ($\nu>1$) or to an oscillatory suppression ($\nu<1$) of the density contrast, which is incompatible with the observed $f\sigma_8$. Consequently, the inclusion of the growth data forces $\nu$ to unity with a precision of $\sigma_\nu\simeq5.2\times10^{-7}$, and all three models collapse to $\Lambda$CDM, with $\Gamma\simeq0.29$--$0.30$ and $H_0\simeq68.8$--$69.1~{\rm km\,s^{-1}\,Mpc^{-1}}$.

    \item \textbf{Bayesian evidence with growth data.} Once $f\sigma_8$ is included, all three models are decisively disfavoured with respect to $\Lambda$CDM, $\Delta\ln Z\simeq13.2$--$13.9$, for every SNe sample. This is not due to a worse fit, since $|\Delta\chi^2|\lesssim0.7$, but entirely to the Occam penalty associated with the extremely narrow posterior of $\nu$ compared with its prior range, as confirmed by both the Laplace approximation and the Savage--Dickey density ratio.
\end{itemize}

These results show that the preference for $\nu\neq1$ found at the background level is not supported once the growth of structure is taken into account: the background expansion of these models remains viable, but the additional freedom introduced by the non-linear matter coupling is not required by the combined data. This highlights the importance of confronting modified gravity scenarios with perturbative observables, since background data alone can favour extensions that are excluded at the level of structure formation.

Several directions deserve further investigation. A complete assessment of the model requires its implementation in a Boltzmann code, enabling a consistent comparison with CMB temperature and polarisation spectra, CMB lensing, and the integrated Sachs--Wolfe effect. Weak-lensing and galaxy-clustering data would provide independent tests of the scale-dependent growth predicted by the model, while relaxing the quasi-static approximation would allow a more accurate treatment of perturbations near the horizon scale. Finally, extensions involving explicit non-minimal couplings between $Q$ and $\mathcal{L}_m$ may modify the effective pressure term and could reconcile a departure from $\Lambda$CDM with the growth data; we leave these questions for future work.

\section*{Acknowledgments}
We acknowledge the computing facilities provided by the ICT Department of Presidency University, Kolkata. RKD thanks Dr.~Supriya Pan and Prof.~Subir Ghosh for helpful discussions.

\clearpage
\appendix
\section*{Supplementary Materials}
\addcontentsline{toc}{section}{APPENDIX}

\begin{table}[t!]
\begin{center}
\resizebox{0.48\textwidth}{!}{%
    \huge
\renewcommand{\arraystretch}{0.1} 
\begin{tabular}{ | @{\hspace{0.1cm}} l @{\hspace{0.4 cm}} l @{\hspace{0.4 cm}} l @{\hspace{0.3 cm}} c @{\hspace{0.3 cm}} c @{\hspace{0.1 cm}} |@{\hspace{0.3cm}} l @{\hspace{0.3 cm}} l @{\hspace{0.4 cm}} l @{\hspace{0.4 cm}} c @{\hspace{0.3 cm}} c @{\hspace{0.1 cm}}|}
\multicolumn{10}{c}{{}}\\
\hline 
\hline
~$z$ & $H(z)$  & $\sigma_{H(z)}$ & Method & Ref. & $z$ & $H(z)$  & $\sigma_{H(z)}$ & Method & Ref. \\
\hline
~0.07  & 69.0  & 19.6  & DA & \cite{Zhang:2012mp} & 0.51  & 90.4  & 1.9   & Clustering & \cite{BOSS:2016wmc} \\
~0.1   & 69.0  & 12.0  & DA  & \cite{Stern:2009ep} & 0.52  & 94.35 & 2.65  & Clustering & \cite{BOSS:2016zkm} \\
~0.12  & 68.6  & 26.2  & DA & \cite{Zhang:2012mp} & 0.56  & 93.33 & 2.32  & Clustering & \cite{BOSS:2016zkm} \\
~0.17  & 83.0  & 8.0   & DA & \cite{Stern:2009ep} & 0.57  & 92.9  & 7.8   & Clustering & \cite{BOSS:2013kxl} \\
~0.1797 & 75.0  & 4.0   & DA & \cite{Moresco:2012jh} & 0.59  & 98.48 & 3.19  & Clustering & \cite{BOSS:2016zkm} \\
~0.1993 & 75.0  & 5.0   & DA & \cite{Moresco:2012jh} & 0.5929 & 104.0 & 13.0  & DA & \cite{Moresco:2012jh} \\
~0.20  & 72.9  & 29.6  & DA  & \cite{Zhang:2012mp} & 0.6   & 87.9  & 6.1   & Clustering & \cite{Blake:2012pj} \\
~0.24  & 79.69 & 2.65  & Clustering & \cite{Gaztanaga:2008xz} & 0.61  & 97.3  & 2.1   & Clustering & \cite{BOSS:2016wmc} \\
~0.27  & 77.0  & 14.0  & DA & \cite{Stern:2009ep} & 0.64  & 98.82 & 2.99  & Clustering & \cite{BOSS:2016zkm} \\
~0.28  & 88.8  & 36.6  & DA & \cite{Zhang:2012mp} & 0.6797 & 92.0  & 8.0   & DA & \cite{Moresco:2012jh} \\
~0.3   & 81.7  & 6.22  & Clustering & \cite{Oka:2013cba} & 0.73  & 97.3  & 7.0   & Clustering & \cite{Blake:2012pj} \\
~0.31  & 78.17 & 4.74  & Clustering & \cite{BOSS:2016zkm} & 0.75  & 98.8  & 33.6  & Clustering & \cite{Borghi:2021rft} \\
~0.34  & 83.8  & 3.66  & Clustering & \cite{Gaztanaga:2008xz} & 0.7812 & 105.0 & 12.0  & DA & \cite{Moresco:2012jh} \\
~0.35  & 82.7  & 8.4   & Clustering & \cite{Chuang:2012qt} & 0.8754 & 125.0 & 17.0  & DA & \cite{Moresco:2012jh} \\
~0.3519 & 83.0  & 14.0  & DA & \cite{Moresco:2012jh} & 0.88  & 90.0  & 40.0  & DA & \cite{Stern:2009ep} \\
~0.36  & 79.93 & 3.39  & Clustering & \cite{BOSS:2016zkm} & 0.9   & 117.0 & 23.0  & DA & \cite{Stern:2009ep} \\
~0.38  & 81.5  & 1.9   & Clustering & \cite{BOSS:2016wmc} & 0.978 & 113.72 & 14.63 & Clustering & \cite{eBOSS:2018yfg} \\
~0.3802 & 83.0  & 13.5  & DA & \cite{Moresco:2016mzx} & 1.037 & 154.0 & 20.0  & DA & \cite{Moresco:2012jh} \\
~0.40  & 82.04 & 2.03  & DA & \cite{BOSS:2016zkm} & 1.23  & 131.44 & 12.42 & Clustering & \cite{eBOSS:2018yfg} \\
~0.4   & 95.0  & 17.0  & DA & \cite{Stern:2009ep} & 1.3   & 168.0 & 17.0  & DA & \cite{Stern:2009ep} \\
~0.4004 & 77.0  & 10.2  & DA & \cite{Moresco:2016mzx} & 1.363 & 160.0 & 33.6  & DA & \cite{Moresco:2015cya} \\
~0.4247 & 87.1  & 11.2  & DA & \cite{Moresco:2016mzx} & 1.43  & 177.0 & 18.0  & DA & \cite{Stern:2009ep} \\
~0.4293 & 91.8  & 5.3   & DA & \cite{Moresco:2016mzx} & 1.526 & 148.11 & 12.71 & Clustering & \cite{eBOSS:2018yfg} \\
~0.43  & 86.45 & 3.68  & Clustering & \cite{Gaztanaga:2008xz} & 1.53  & 140.0 & 14.0  & DA & \cite{Stern:2009ep} \\
~0.44  & 82.6  & 7.8   & Clustering & \cite{Blake:2012pj} & 1.75  & 202.0 & 40.0  & DA & \cite{Stern:2009ep} \\
~0.44  & 84.81 & 1.83  & Clustering & \cite{BOSS:2016zkm} & 1.944 & 172.63 & 14.79 & Clustering & \cite{eBOSS:2018yfg} \\
~0.4497 & 92.8  & 12.9  & DA & \cite{Moresco:2016mzx} & 1.965 & 186.5 & 50.4  & DA & \cite{Moresco:2015cya} \\
~0.47  & 89.0  & 34.0  & DA & \cite{Ratsimbazafy_2017} & 2.3   & 224.0 & 8.0   & Clustering & \cite{BOSS:2014hwf} \\
~0.4783 & 80.9  & 9.0   & DA & \cite{Moresco:2016mzx} & 2.33  & 224.0 & 8.0   & Clustering & \cite{BOSS:2017fdr} \\
~0.48  & 87.79 & 2.03  & DA & \cite{BOSS:2016zkm} & 2.34  & 222.0 & 7.0   & Clustering & \cite{BOSS:2014hwf} \\
~0.48  & 97.0  & 62.0  & DA & \cite{Stern:2009ep} & 2.36  & 226.0 & 8.0   & Clustering & \cite{BOSS:2013igd} \\
\hline

\end{tabular}}
\end{center}
\caption{The complete set of 62 Observational Hubble Data (OHD), along with their associated uncertainties, utilized in this work. Here, {\bf DA} refers to the differential age (Cosmic Chronometer) method, while {\bf clustering} corresponds to measurements from radial BAO surveys. }
\label{table_OHD_data}
\end{table}

\begin{table}
\centering
 \resizebox{1.0\columnwidth}{!}{
 \large
\renewcommand{\arraystretch}{1.0}
\begin{tabular}{|l c c c|}
\hline
{ \textbf{Dataset}} & {\textbf{$z$} } & {  $\boldmath{f\sigma_8(z)}$} & {\textbf{Ref.}}\\
\hline\hline 
~2MTF~ & $0.001$ & $0.505 \pm 0.085$ & \cite{Howlett:2017asq} \\
~6dFGS+SNIa~ & $0.020$ & $0.4280 \pm 0.0465$ & \cite{Huterer:2016uyq}\\
~IRAS+SNIa~ & $0.020$ & $0.398 \pm 0.065$ & \cite{Hudson:2012gt, Turnbull:2011ty}\\
~2MASS~ & $0.020$ & $0.314 \pm 0.048$ & \cite{Hudson:2012gt, Davis:2010sw}\\
~2dFGRS~ & $0.170$ & $0.510 \pm 0.060$ & \cite{Song:2008qt}\\
~GAMA~ & $0.180$ & $0.360 \pm 0.090$ & \cite{Blake:2013nif}\\
~GAMA~ & $0.380$ & $0.440 \pm 0.060$ & \cite{Blake:2013nif}\\
~SDSS-IV (MGS)~ & $0.150$ & $0.530 \pm 0.160$ & \cite{eBOSS:2020yzd} \\
~SDSS-IV (BOSS Galaxy)~ & $0.380$ & $0.497 \pm 0.045$ & \cite{eBOSS:2020yzd}\\
~SDSS-IV (BOSS Galaxy)~ & $0.510$ & $0.459 \pm 0.038$ & \cite{eBOSS:2020yzd}\\
~SDSS-IV (eBOSS LRG)~ & $0.700$ & $0.473 \pm 0.041$ & \cite{eBOSS:2020yzd}\\
~SDSS-IV (eBOSS ELG)~ & $0.850$ & $0.315 \pm 0.095$ & \cite{eBOSS:2020yzd}\\
~WiggleZ~ & $0.44$ & $0.413 \pm 0.080$ & \cite{Blake:2012pj} \\
~WiggleZ~ & $0.600$ & $0.390 \pm 0.063$ & \cite{Blake:2012pj}\\
~WiggleZ~ & $0.730$ & $0.437 \pm 0.072$ & \cite{Blake:2012pj}\\
~Vipers PDR-2~ & $0.600$ & $0.550 \pm 0.120$ & \cite{Pezzotta:2016gbo} \\
~Vipers PDR-2~ & $0.860$ & $0.400 \pm 0.110$ & \cite{Pezzotta:2016gbo} \\
~FastSound~ & $1.400$ & $0.482 \pm 0.116$ & \cite{Okumura:2015lvp} \\
~SDSS-IV (eBOSS Quasar)~ & $0.978$ & $0.379 \pm 0.176$ & \cite{eBOSS:2018yfg}\\
~SDSS-IV (eBOSS Quasar)~ & $1.230$ & $0.385 \pm 0.099$ & \cite{eBOSS:2018yfg}\\
~SDSS-IV (eBOSS Quasar)~ & $1.526$ & $0.342 \pm 0.070$ & \cite{eBOSS:2018yfg}\\
~SDSS-IV (eBOSS Quasar)~ & $1.944$ & $0.364 \pm 0.106$ & \cite{eBOSS:2018yfg}\\ 
\hline
\end{tabular}}
\caption{$f\sigma_8(z)$ measurements obtained from different observational surveys, which have been used in this work.}
 \label{tab:fsigma8}
\end{table} 

\section{Observational Data and Methodology Overview}

In this section, we describe the observational probes employed from multiple cosmological surveys and outline the statistical inference framework used to constrain the parameters of the models introduced in Sec. \ref{sec:mod_theory}. The observational probes are outlined as follows:

\begin{itemize}
    \item[1.] \textbf{Observational Hubble Data (OHD):} Measurements of the Hubble parameter $H(z)$ in the low-redshift regime are obtained using two complementary approaches. The first relies on the differential age (DA) of passively evolving galaxies at different redshifts, commonly referred to as the cosmic chronometer method, while the second exploits the radial, line-of-sight scale of baryon acoustic oscillation (BAO) features, known as the radial BAO method. In this letter, we employ a total of 62 OHD measurements \cite{Yang:2023qsz}, consisting of 34 data points derived from the DA technique and 28 from the radial BAO method, as summarized in Table~\ref{table_OHD_data}.

    \item[2.] \textbf{Baryon Acoustic Oscillations (BAO):} We utilize recent BAO measurements from the first two years of observations of the Dark Energy Spectroscopic Instrument (DESI), covering the redshift range $0.0295 \le z \le 2.330$ and divided into nine effective redshift bins, as summarized in Table~IV of Ref.~\cite{DESI:2025zgx}. In this dataset, the BAO feature in the matter clustering is treated as a standard ruler. The isotropic BAO measurements are expressed in terms of the volume-averaged distance $D_V(z)/r_d$, while the anisotropic BAO measurements provide constraints on the comoving angular diameter distance $D_M(z)/r_d$ and the Hubble distance $D_H(z)/r_d$, where $r_d$ denotes the comoving sound horizon at the drag epoch. For brevity, we refer to this dataset as \texttt{DESI}. 
    
    Since our analysis is restricted to the late-time background evolution, the sound horizon is calibrated using the Planck $\Lambda$CDM cosmology, with a Gaussian prior of $r_d = 147.09 \pm 0.26$~Mpc \cite{Planck:2018vyg}.

    \item[3.] \textbf{Type Ia Supernova (SNIa): } Type Ia supernovae (SNIa) are established standard candles and provide a key probe of the late-time accelerated expansion of the Universe. Their utility arises from their nearly uniform intrinsic luminosity, allowing robust determinations of the luminosity distance–redshift relation. In our analysis, we incorporated three SNIa datasets, as follows: 
    
    \begin{enumerate}
        \item[(i)] \textbf{Pantheon$+$:} We use a compilation of 1701 light-curve measurements corresponding to 1550 distinct Type Ia supernovae spanning the redshift range $z \in [0.001, 2.26]$~\cite{Scolnic:2021amr, Brout:2022vxf}. In this work, we consider a subsample of 1590 SNIa within the interval $0.01016 \leq z \leq 2.26137$ to mitigate the impact of peculiar velocity corrections~\cite{Chan-GyungPark:2024mlx}. We refer to dataset as \texttt{\bf PP}.
        \item[(ii)] \textbf{Union3:} The latest Union3 compilation consists of 2087 SNIa drawn from 24 independent datasets, homogenized onto a common distance scale using the SALT3 light-curve fitter. The data are analyzed within the UNITY1.5 Bayesian framework and compressed into 22 binned distance modulus measurements over the redshift range $0.05 < z < 2.26$, which we adopt in this work~\cite{Rubin:2023ovl}. We denote this subsample as \texttt{\bf U3}.
        \item[(iii)] \textbf{DESY5:} We employ the latest release from the Dark Energy Survey Supernova Program (DES-Dovekie), comprising 1820 photometric Type Ia supernova light curves~\cite{DES:2025sig}. This sample includes 1623 DES SNIa in the redshift interval $0.025 < z < 1.145$, along with 197 low-redshift ($z < 0.1$) SNIa from the CfA3~\cite{Hicken:2009df}, CfA4~\cite{Hicken:2012zr}, CSP~\cite{Krisciunas:2017yoe}, and Foundation~\cite{Foley:2017zdq} surveys. Throughout this work, we refer to this dataset as \texttt{\bf DESY5}.
    \end{enumerate}

    \item[4.] \textbf{Redshift-Space Distortions (RSD):} 
    Redshift-space distortions provide a direct observational probe of the growth of large-scale structure. Galaxy peculiar velocities, arising from gravitational infall, induce anisotropies in the observed clustering pattern along the line of sight. Since these anisotropies are governed by the growth rate of matter perturbations, RSD measurements offer a powerful test of gravitational physics on cosmological scales and enable constraints on possible deviations from general relativity. In this work, we incorporate RSD observations through the combined observable $f\sigma_8(a)$, where $f(a)$ denotes the linear growth rate and $\sigma_8(a)$ represents the rms amplitude of matter fluctuations. The 22 uncorrelated $f\sigma_8(a)$ data points used in our analysis are summarized in Table~\ref{tab:fsigma8}.
\end{itemize}

To obtain robust constraints on the model parameters from the combined observational datasets, we perform a comprehensive Markov Chain Monte Carlo (MCMC) analysis using the affine-invariant ensemble sampler implemented in the publicly available \textsc{emcee} package~\cite{Foreman-Mackey:2012any}\footnote{\href{https://github.com/dfm/emcee}{https://github.com/dfm/emcee}}. We assume identical uniform priors for all three models, specified as $H_0 \in [50,100]$, $\Gamma \in [0,1]$, and $\nu \in [0.6,2.0]$. The resulting MCMC chains are post-processed and visualized using the \textsc{GetDist} package~\cite{Lewis:2019xzd}\footnote{\href{https://github.com/cmbant/getdist}{https://github.com/cmbant/getdist}}.\\

\vspace{1cm}
\section{Non-conservation}
\subsection{Re-arrangement of field equation}
We begin by recalling the field equations in the context of $f(Q,\mathcal{L}_m)$ gravity, which are given by:
\begin{align*}
 \frac{2}{\sqrt{-g}}\nabla_{\alpha}(f_Q \sqrt{-g} P^{\alpha}_{\rho\nu}) + f_Q (P_{\mu\alpha\beta} Q_{\nu}^{\alpha\beta} - 2 Q^{\alpha\beta}{}_{\mu} P_{\alpha\beta\nu}) \nonumber \\ + \frac{f}{2} g_{\rho\nu} = \frac{f_{\mathcal{L}_m}}{2} (g_{\rho\nu} \mathcal{L}_m - T_{\rho\nu}) 
\end{align*}
To obtain the field equations in mixed tensor form, we raise the free index by multiplying both sides of the equation by the inverse metric $g^{\mu\rho}$. Applying this algebraic operation to the first term containing the covariant derivative yields:
\begin{align*}   
 g^{\mu\rho} \left[ \frac{2}{\sqrt{-g}}\nabla_{\alpha}(f_Q\sqrt{-g})P^{\alpha}_{\rho\nu} + \frac{2}{\sqrt{-g}}f_Q\sqrt{-g}\nabla_{\alpha}P^{\alpha}_{\rho\nu} \right] \\
= \frac{2}{\sqrt{-g}} \nabla_{\alpha}(f_Q\sqrt{-g}) g^{\mu\rho} P^{\alpha}_{\rho\nu} + 2f_Q g^{\mu\rho}\nabla_{\alpha}P^{\alpha}_{\rho\nu} 
\end{align*}

To evaluate the second term in the bracket, we utilize the product rule for the covariant derivative:
$ g^{\mu\rho}\nabla_{\alpha}P^{\alpha}_{\rho\nu} = \nabla_{\alpha}(g^{\mu\rho}P^{\alpha}_{\rho\nu}) - (\nabla_{\alpha}g^{\mu\rho})P^{\alpha}_{\rho\nu} $

Furthermore, recalling the definition of the non-metricity tensor \cite{Xu:2019sbp}, we have:
$ \nabla_{\alpha}g^{\mu\rho} = -Q_{\alpha}^{\phantom{\alpha}\rho\mu} $ 

Substituting this relation back, the expression reformulates as:
$ g^{\mu\rho}\nabla_{\alpha}P^{\alpha}_{\rho\nu} = \nabla_{\alpha}P^{\alpha\mu}_{\phantom{\alpha\mu}\nu} + Q_{\alpha}^{\phantom{\alpha}\rho\mu}P^{\alpha}_{\rho\nu} $

Incorporating this identity into the expanded derivative term, we obtain:
\begin{align*}
 & g^{\mu\rho} \left[ \frac{2}{\sqrt{-g}}\nabla_{\alpha}(f_Q\sqrt{-g})P^{\alpha}_{\rho\nu} + \frac{2}{\sqrt{-g}}f_Q\sqrt{-g}\nabla_{\alpha}P^{\alpha}_{\rho\nu} \right]\\  = &\frac{2}{\sqrt{-g}} \left[ \nabla_{\alpha}(f_Q\sqrt{-g})P^{\alpha\mu}_{\phantom{\alpha\mu}\nu} + f_Q\sqrt{-g}\nabla_{\alpha}P^{\alpha\mu}_{\phantom{\alpha\mu}\nu} \right] + 2f_Q Q_{\alpha}^{\phantom{\alpha}\rho\mu}P^{\alpha}_{\rho\nu} \\
 = &\frac{2}{\sqrt{-g}}\nabla_{\alpha}(f_Q\sqrt{-g}P^{\alpha\mu}_{\phantom{\alpha\mu}\nu}) + 2f_Q Q_{\alpha}^{\phantom{\alpha}\rho\mu}P^{\alpha}_{\rho\nu} 
\end{align*}
Next, we manipulate the indices of the residual non-metricity term using the metric tensor:
$ Q_{\alpha}^{\phantom{\alpha}\rho\mu} = g_{\alpha\kappa} Q^{\kappa\rho\mu}$ and $ 
 P^{\alpha}_{\rho\nu} = g^{\alpha z} P_{z\rho\nu}$  

Applying these relations, the residual term simplifies via the Kronecker delta: $ 
 2f_Q Q_{\alpha}^{\phantom{\alpha}\rho\mu}P^{\alpha}_{\rho\nu} = 2f_Q g_{\alpha\kappa}g^{\alpha z} Q^{\kappa\rho\mu}P_{z\rho\nu} 
 = 2f_Q \delta_{\kappa}^{z} Q^{\kappa\rho\mu}P_{z\rho\nu} 
 = 2f_Q Q^{\alpha\beta\mu}P_{\alpha\beta\nu} $. 
 \begin{widetext}

Consequently, the complete first term of the field equation can be written as:
\begin{align*}
 \frac{2}{\sqrt{-g}}\nabla_{\alpha}(f_Q\sqrt{-g}P^{\alpha\mu}_{\phantom{\alpha\mu}\nu}) + 2f_Q Q_{\alpha}^{\phantom{\alpha}\rho\mu}P^{\alpha}_{\rho\nu} = \frac{2}{\sqrt{-g}}\nabla_{\alpha}(f_Q\sqrt{-g}P^{\alpha\mu}_{\phantom{\alpha\mu}\nu}) + 2f_Q Q^{\alpha\beta\mu}P_{\alpha\beta\nu} 
\end{align*}

We now proceed to evaluate the remaining terms on the left-hand side by applying the inverse metric $g^{\mu\rho}$:

\begin{align*}   
 g^{\mu\rho} \left\{ f_Q \left[ (P_{\rho\alpha\beta} Q_{\nu}^{\alpha\beta}) - 2 Q^{\alpha\beta}_{\phantom{\alpha\beta}\rho} P_{\alpha\beta\nu} \right] \right\} =  f_Q (P^{\mu}_{\phantom{\mu}\alpha\beta} Q_{\nu}^{\alpha\beta} - 2 Q^{\alpha\beta\mu} P_{\alpha\beta\nu}) 
\end{align*}

Similarly, contracting the metric tensor yields the Kronecker delta, $ \frac{f}{2} g^{\mu\rho} g_{\rho\nu} = \frac{f}{2} \delta_{\nu}^{\mu} $, while the matter sector on the right-hand side trivially evaluates to: 
$\frac{f_{\mathcal{L}_m}}{2} g^{\mu\rho} (g_{\rho\nu} \mathcal{L}_m - T_{\rho\nu}) = \frac{f_{\mathcal{L}_m}}{2} (\delta_{\nu}^{\mu} \mathcal{L}_m - T_{\nu}^{\mu})$.

Assembling all evaluated components, we find that the residual non-metricity terms cancel out exactly:
\begin{align*}
& \frac{2}{\sqrt{-g}}\nabla_{\alpha}(f_Q\sqrt{-g}P^{\alpha\mu}_{\phantom{\alpha\mu}\nu}) + \bcancel{2f_Q Q^{\alpha\beta\mu}P_{\alpha\beta\nu}} + f_Q (P^{\mu}_{\phantom{\mu}\alpha\beta} Q_{\nu}^{\alpha\beta}) - \bcancel{2f_Q Q^{\alpha\beta\mu}P_{\alpha\beta\nu}} + \frac{f}{2}\delta_{\nu}^{\mu} = \frac{f_{\mathcal{L}_m}}{2}(\delta_{\nu}^{\mu} \mathcal{L}_m - T_{\nu}^{\mu}) 
\end{align*}
Therefore, the final field equation in mixed tensor representation reduces cleanly to: 
\begin{align}
& \frac{2}{\sqrt{-g}}\nabla_{\alpha}(f_Q\sqrt{-g}P^{\alpha\mu}_{\phantom{\alpha\mu}\nu}) + f_Q P^{\mu}_{\phantom{\mu}\alpha\beta} Q_{\nu}^{\alpha\beta} + \frac{f}{2}\delta_{\nu}^{\mu} = \frac{f_{\mathcal{L}_m}}{2}(\delta_{\nu}^{\mu} \mathcal{L}_m - T_{\nu}^{\mu})
\label{re-field}
\end{align}

\subsection{Relationship Between Affine and Levi-Civita Connections and proceeding non-conservation}

To fully evaluate the divergence of the energy-momentum tensor and transition smoothly between geometries, we must explicitly define the transformation between the general affine covariant derivative ($\nabla_\mu$) and the Levi-Civita covariant derivative ($D_\mu$).

As outlined in the derivation, the affine derivative of an arbitary  mixed tensor $K^{\mu}_{\nu}$ relates to its Levi-Civita counterpart by subtracting the non-metricity and deformation contributions:

\begin{equation}
D_{\mu} K^{\mu}_{\nu} = \nabla_{\mu} K^{\mu}_{\nu} + \frac{1}{2} Q_{\mu} K^{\mu}_{\nu} + L^{\rho}_{\mu\nu} K^{\mu}_{\rho}
\end{equation}

From this contraction, we can abstract the general operator equivalence that bridges the two spaces. The Levi-Civita derivative operator expands as:
\begin{equation}
D_{\mu} = \nabla_{\mu} + \frac{1}{2} Q_{\mu} + L^{\rho}_{\mu\nu}
\end{equation}

Here, we identify the specific geometric components governing the transformation:
\begin{itemize}
    \item $D_{\mu}$ corresponds to the \textbf{Levi-Civita connection}.
    \item $\nabla_{\mu}$ corresponds to the general \textbf{Affine connection}.
    \item $\frac{1}{2} Q_{\mu}$ represents the \textbf{non-metricity} vector trace contribution.
    \item $L^{\rho}_{\mu\nu}$ represents the \textbf{disformation}  tensor.
\end{itemize}

This operator relation is the mathematical tool that allows us to expand the $D_\mu$ derivatives in the $f(Q, \mathcal{L}_m)$ field equation. This treatment is required to establish the non-conservation equation. The regarding detailed descriptions are available in ~\cite{Xu:2019sbp,Harko:2018gxr}. At first we imply it on right hand sided part of equation ~\eqref{re-field} and we get

\begin{align}
D_{\mu}\left\{\frac{1}{2}f_{\mathcal{L}_m}(\delta_{\nu}^{\mu}\mathcal{L}_m-T_{\nu}^{\mu})\right\} &= D_{\mu}\left(\frac{f_{\mathcal{L}_m}}{2}\delta_{\nu}^{\mu}\mathcal{L}_m\right) - \left(D_{\mu}T_{\nu}^{\mu}\right)\frac{f_{\mathcal{L}_m}}{2} - T_{\nu}^{\mu}D_{\mu}\left(\frac{f_{\mathcal{L}_m}}{2}\right) \nonumber \\
&= \delta_{\nu}^{\mu}D_{\mu}\left(\frac{f_{\mathcal{L}_m}}{2}\mathcal{L}_m\right) + \frac{f_{\mathcal{L}_m}\mathcal{L}_m}{2}D_{\mu}(\delta_{\nu}^{\mu}) - \frac{f_{\mathcal{L}_m}}{2}(D_{\mu}T_{\nu}^{\mu}) - T_{\nu}^{\mu}D_{\mu}\left(\frac{f_{\mathcal{L}_m}}{2}\right) \nonumber \\
&= \frac{1}{2}f_{\mathcal{L}_m}\partial_{\nu}\mathcal{L}_m + \frac{1}{2}\mathcal{L}_m\partial_{\nu}f_{\mathcal{L}_m} - \frac{1}{2}T_{\nu}^{\mu}\partial_{\mu}f_{\mathcal{L}_m} - \frac{f_{\mathcal{L}_m}}{2}D_{\mu}T_{\nu}^{\mu}
\end{align}

Now we will deal with the left hand side  part as it will look like;

\begin{equation}
LHS = \frac{2}{\sqrt{-g}}\nabla_{\alpha}(f_Q\sqrt{-g}P^{\alpha\mu}_{\phantom{\alpha\mu}\nu}) + f_Q P^{\mu}_{\phantom{\mu}\alpha\beta}Q_{\nu}^{\alpha\beta} + \delta_{\nu}^{\mu}\left(\frac{f}{2}\right)
\end{equation}
So we can imply the covariant derivative in the following scenario;
\begin{equation}
D_{\mu}(LHS) = D_{\mu}(LHS_1) + D_{\mu}(LHS_2) + D_{\mu}(LHS_3)
\end{equation}
If we try to break and rearrange the term of 1st part will represnt as;
\begin{align}
D_{\mu}(LHS_1) &= D_{\mu}\left( \frac{2}{\sqrt{-g}}\nabla_{\alpha}(f_Q\sqrt{-g}P^{\alpha\mu}_{\phantom{\alpha\mu}\nu}) \right) \nonumber \\
&= \frac{2}{\sqrt{-g}} \Bigg[ \nabla_{\mu} \nabla_{\alpha} (f_Q \sqrt{-g} P^{\alpha\mu}_{\phantom{\alpha\mu}\nu}) + \frac{1}{2} Q_{\mu} \nabla_{\alpha} (f_Q \sqrt{-g} P^{\alpha\mu}_{\phantom{\alpha\mu}\nu}) \nonumber \\
&\quad + L^{\rho}_{\mu\nu} \Big( f_Q P^{\alpha\mu}_{\phantom{\alpha\mu}\rho} (\nabla_{\alpha}\sqrt{-g}) + \sqrt{-g} P^{\alpha\mu}_{\phantom{\alpha\mu}\rho} (\nabla_{\alpha} f_Q) + \sqrt{-g} f_Q (\nabla_{\alpha} P^{\alpha\mu}_{\phantom{\alpha\mu}\rho}) \Big) \Bigg] \nonumber \\
&= \frac{2}{\sqrt{-g}} \nabla_{\mu} \nabla_{\alpha} (f_Q \sqrt{-g} P^{\alpha\mu}_{\phantom{\alpha\mu}\nu}) + \frac{Q_{\mu}}{\sqrt{-g}} \nabla_{\alpha} (f_Q \sqrt{-g} P^{\alpha\mu}_{\phantom{\alpha\mu}\nu}) \nonumber \\
&\quad + \frac{2}{\sqrt{-g}} L^{\rho}_{\mu\nu} \left( f_Q P^{\alpha\mu}_{\phantom{\alpha\mu}\rho} \frac{\sqrt{-g}}{2} Q_{\alpha} + \sqrt{-g} P^{\alpha\mu}_{\phantom{\alpha\mu}\rho} (\nabla_{\alpha} f_Q) + \sqrt{-g} f_Q (\nabla_{\alpha} P^{\alpha\mu}_{\phantom{\alpha\mu}\rho}) \right) \nonumber \\
&= \frac{2}{\sqrt{-g}} \nabla_{\mu} \nabla_{\alpha} (f_Q \sqrt{-g} P^{\alpha\mu}_{\phantom{\alpha\mu}\nu}) + \frac{Q_{\mu}}{\sqrt{-g}} \nabla_{\alpha} (f_Q \sqrt{-g} P^{\alpha\mu}_{\phantom{\alpha\mu}\nu}) \nonumber \\
&\quad + L^{\rho}_{\mu\nu} Q_{\alpha} f_Q P^{\alpha\mu}_{\phantom{\alpha\mu}\rho} + 2 P^{\alpha\mu}_{\phantom{\alpha\mu}\rho} L^{\rho}_{\mu\nu} (\nabla_{\alpha} f_Q) + 2 L^{\rho}_{\mu\nu} f_Q (\nabla_{\alpha} P^{\alpha\mu}_{\phantom{\alpha\mu}\rho})
\end{align}
As $\sqrt{-g}$ is a scalar density so it will not take participation in covariant derivative operation and so the other second part 
\begin{align}
D_{\mu}(LHS_2) &= \nabla_{\mu}(f_Q P^{\mu\alpha\beta}Q_{\nu\alpha\beta}) + \frac{1}{2}Q_{\mu}f_Q P^{\mu\alpha\beta}Q_{\nu\alpha\beta} + L_{\mu\nu}^{\rho}f_Q P^{\mu\alpha\beta}Q_{\rho\alpha\beta} \nonumber \\
&= (\nabla_{\mu}f_Q)P^{\mu\alpha\beta}Q_{\nu\alpha\beta} + f_Q(\nabla_{\mu}P^{\mu\alpha\beta})Q_{\nu\alpha\beta} + f_Q P^{\mu\alpha\beta}(\nabla_{\mu}Q_{\nu\alpha\beta}) \nonumber \\
&\quad + \frac{1}{2}Q_{\mu}f_Q P^{\mu\alpha\beta}Q_{\nu\alpha\beta} + L_{\mu\nu}^{\rho}f_Q P^{\mu\alpha\beta}Q_{\rho\alpha\beta}
\end{align}
The remainig third part is quite easy to handel and it will be,
\begin{equation}
D_{\mu}(LHS_3) = D_{\mu}\left(\frac{\delta_{\nu}^{\mu}f}{2}\right) = \frac{1}{2}(f_Q\partial_{\nu}Q + f_{\mathcal{L}_m}\partial_{\nu}\mathcal{L}_m)
\end{equation}

\begin{align}
D_{\mu}(\mathrm{RHS}) &= \frac{2}{\sqrt{-g}} \nabla_{\mu} \nabla_{\alpha} (f_Q \sqrt{-g} P^{\alpha\mu}_{\phantom{\alpha\mu}\nu}) \quad \circled{1} \nonumber \\
&\quad + \frac{Q_{\mu}}{\sqrt{-g}} \nabla_{\alpha} (f_Q \sqrt{-g} P^{\alpha\mu}_{\phantom{\alpha\mu}\nu}) \quad \circled{2} \nonumber \\
&\quad + L^{\rho}_{\phantom{\rho}\mu\nu} Q_{\alpha} f_Q P^{\alpha\mu}_{\phantom{\alpha\mu}\rho} \quad \circled{3} \nonumber \\
&\quad + 2 P^{\alpha\mu}_{\phantom{\alpha\mu}\rho} L^{\rho}_{\phantom{\rho}\mu\nu} (\nabla_{\alpha} f_Q) \quad \circled{4} \nonumber \\
&\quad + 2 L^{\rho}_{\phantom{\rho}\mu\nu} f_Q (\nabla_{\alpha} P^{\alpha\mu}_{\phantom{\alpha\mu}\rho}) \quad \circled{5} \nonumber \\
&\quad + (\nabla_{\mu} f_Q) P^{\mu\alpha\beta} Q_{\nu\alpha\beta} \quad \circled{6} \nonumber \\
&\quad + f_Q (\nabla_{\mu} P^{\mu\alpha\beta}) Q_{\nu\alpha\beta} \quad \circled{7} \nonumber \\
&\quad + f_Q P^{\mu\alpha\beta} (\nabla_{\mu} Q_{\nu\alpha\beta}) \quad \circled{8} \nonumber \\
&\quad + \frac{1}{2} Q_{\mu} f_Q P^{\mu\alpha\beta} Q_{\nu\alpha\beta} \quad \circled{9} \nonumber \\
&\quad + L^{\rho}_{\phantom{\rho}\mu\nu} f_Q P^{\mu\alpha\beta} Q_{\rho\alpha\beta} \quad \circled{10} \nonumber \\
&\quad + \frac{f_Q \partial_{\nu} Q + f_{\mathcal{L}_m} \partial_{\nu} \mathcal{L}_m}{2} \quad \circled{11}
\end{align}

Using the connection field equation,
\begin{equation}
4 \nabla_{\mu} \nabla_{\nu} (f_Q \sqrt{-g} P^{\mu\nu}_{\phantom{\mu\nu}\alpha}) = - \nabla_{\mu} \nabla_{\nu} H_{\alpha}^{\phantom{\alpha}\mu\nu},
\end{equation}
and the fact that in symmetric teleparallel geometry ($R^{\alpha}_{\phantom{\alpha}\beta\mu\nu}=0$, $T^{\alpha}_{\phantom{\alpha}\mu\nu}=0$) covariant derivatives commute, $\nabla_{\mu}\nabla_{\alpha}=\nabla_{\alpha}\nabla_{\mu}$, the first term becomes
\begin{equation}
\circled{1} = \frac{2}{\sqrt{-g}} \nabla_{\alpha} \nabla_{\mu} (f_Q \sqrt{-g} P^{\alpha\mu}_{\phantom{\alpha\mu}\nu})
= -\frac{1}{2\sqrt{-g}} \nabla_{\alpha} \nabla_{\mu} H^{\alpha\mu}_{\phantom{\alpha\mu}\nu} .
\end{equation}

The remaining terms combine pairwise, using $(Q_{\nu\alpha\beta} + 2 L_{\beta\alpha\nu}) P^{\mu\alpha\beta} = 0$:
\begin{align}
\circled{3} + \circled{9} &= L^{\rho}_{\phantom{\rho}\mu\nu} Q_{\alpha} f_Q P^{\alpha\mu}_{\phantom{\alpha\mu}\rho} + \frac{1}{2} Q_{\mu} f_Q P^{\mu\alpha\beta} Q_{\nu\alpha\beta} \nonumber \\
&= f_Q Q_{\mu} L_{\beta\alpha\nu} P^{\mu\alpha\beta} + \frac{f_Q Q_{\mu}}{2} P^{\mu\alpha\beta} Q_{\nu\alpha\beta} \nonumber \\
&= \frac{f_Q Q_{\mu}}{2} (Q_{\nu\alpha\beta} + 2 L_{\beta\alpha\nu}) P^{\mu\alpha\beta} = 0 ,
\end{align}

\begin{align}
\circled{4} + \circled{6} &= 2 P^{\alpha\mu}_{\phantom{\alpha\mu}\rho} L^{\rho}_{\phantom{\rho}\mu\nu} (\nabla_{\alpha} f_Q) + (\nabla_{\mu} f_Q) P^{\mu\alpha\beta} Q_{\nu\alpha\beta} \nonumber \\
&= (\nabla_{\mu} f_Q) (Q_{\nu\alpha\beta} + 2 L_{\beta\alpha\nu}) P^{\mu\alpha\beta} = 0 .
\end{align}

In \circled{5} the index of $P^{\alpha\mu}_{\phantom{\alpha\mu}\rho}$ is lowered with the metric, so that
$\nabla_{\alpha} P^{\alpha\mu}_{\phantom{\alpha\mu}\rho} = Q_{\alpha\rho\kappa} P^{\alpha\mu\kappa} + g_{\rho\kappa} \nabla_{\alpha} P^{\alpha\mu\kappa}$, which gives
\begin{align}
\circled{5} + \circled{7} &= 2 L^{\rho}_{\phantom{\rho}\alpha\nu} f_Q Q_{\mu\rho\beta} P^{\mu\alpha\beta}
+ f_Q (Q_{\nu\alpha\beta} + 2 L_{\beta\alpha\nu}) \nabla_{\mu} P^{\mu\alpha\beta} \nonumber \\
&= 2 L^{\rho}_{\phantom{\rho}\alpha\nu} f_Q Q_{\mu\beta\rho} P^{\mu\alpha\beta} .
\end{align}

Adding \circled{8} and \circled{10}, and using $\nabla_{\mu} Q_{\nu\alpha\beta} = \nabla_{\nu} Q_{\mu\alpha\beta}$ together with the symmetry $P^{\mu\alpha\beta}=P^{\mu\beta\alpha}$,
\begin{align}
\circled{5} + \circled{7} + \circled{8} + \circled{10}
&= f_Q \left[ \nabla_{\nu} Q_{\mu\alpha\beta} + L^{\rho}_{\phantom{\rho}\alpha\nu} Q_{\mu\rho\beta} + L^{\rho}_{\phantom{\rho}\beta\nu} Q_{\mu\alpha\rho} + L^{\rho}_{\phantom{\rho}\nu\mu} Q_{\rho\alpha\beta} \right] P^{\mu\alpha\beta} \nonumber \\
&= f_Q P^{\mu\alpha\beta} D_{\nu} Q_{\mu\alpha\beta}
= \frac{1}{2} f_Q D_{\nu} \left( Q_{\mu\alpha\beta} P^{\mu\alpha\beta} \right)
= -\frac{1}{2} f_Q \partial_{\nu} Q ,
\end{align}
so that
\begin{align}
\circled{5} + \circled{7} + \circled{8} + \circled{10} + \circled{11}
&= -\frac{1}{2} f_Q \partial_{\nu} Q + \frac{f_Q \partial_{\nu} Q + f_{\mathcal{L}_m} \partial_{\nu} \mathcal{L}_m}{2}
= \frac{1}{2} f_{\mathcal{L}_m} \partial_{\nu} \mathcal{L}_m .
\end{align}

Substituting all evaluated components and equating to the divergence of the left-hand side,
\begin{align}
&\frac{1}{2}f_{\mathcal{L}_m}\partial_{\nu}\mathcal{L}_m + \frac{1}{2}\mathcal{L}_m\partial_{\nu}f_{\mathcal{L}_m} - \frac{1}{2}T^{\mu}_{\phantom{\mu}\nu}\partial_{\mu}f_{\mathcal{L}_m} - \frac{f_{\mathcal{L}_m}}{2}D_{\mu}T^{\mu}_{\phantom{\mu}\nu} \nonumber \\
&\qquad = -\frac{1}{2\sqrt{-g}}\nabla_{\alpha}\nabla_{\mu}H^{\alpha\mu}_{\phantom{\alpha\mu}\nu} + \frac{Q_{\mu}}{\sqrt{-g}}\nabla_{\alpha}(f_Q\sqrt{-g}P^{\alpha\mu}_{\phantom{\alpha\mu}\nu}) + \frac{1}{2}f_{\mathcal{L}_m}\partial_{\nu}\mathcal{L}_m .
\end{align}
Cancelling the common term and isolating the divergence of the energy-momentum tensor,
\begin{equation}
\frac{1}{2}(\delta^{\mu}_{\nu}\mathcal{L}_m-T^{\mu}_{\phantom{\mu}\nu})\partial_{\mu}f_{\mathcal{L}_m} - \frac{f_{\mathcal{L}_m}}{2}D_{\mu}T^{\mu}_{\phantom{\mu}\nu} = -\frac{1}{2\sqrt{-g}}\nabla_{\alpha}\nabla_{\mu}H^{\alpha\mu}_{\phantom{\alpha\mu}\nu} + \frac{Q_{\mu}}{\sqrt{-g}}\nabla_{\alpha}(f_Q\sqrt{-g}P^{\alpha\mu}_{\phantom{\alpha\mu}\nu}) ,
\end{equation}
and finally
\begin{equation}
D_{\mu}T^{\mu}_{\phantom{\mu}\nu} = \frac{(\delta^{\mu}_{\nu}\mathcal{L}_m - T^{\mu}_{\phantom{\mu}\nu})\,\partial_{\mu}f_{\mathcal{L}_m}}{f_{\mathcal{L}_m}} + \frac{1}{f_{\mathcal{L}_m}\sqrt{-g}} \left[ \nabla_{\alpha}\nabla_{\mu}H^{\alpha\mu}_{\phantom{\alpha\mu}\nu} - 2Q_{\mu}\nabla_{\alpha}(f_Q\sqrt{-g}P^{\alpha\mu}_{\phantom{\alpha\mu}\nu}) \right] .
\end{equation}

\end{widetext}

\section{Perturbation Equations}
As previously mentioned, we adopt the conformal Newtonian gauge. By working in the coincident gauge, the covariant derivative simplifies to the standard partial derivative, yielding $\nabla_{\alpha} \overset{0}{=} \partial_{\alpha}$\cite{Harko:2018gxr,BeltranJimenez:2017tkd}.

To introduce perturbations to the field equation~\eqref{field}, we must first perturb the fundamental geometric quantities as follows:,
\begin{align}
\delta Q_{\alpha\mu\nu} &= \partial_{\alpha}\delta g_{\mu\nu} \\
    \delta Q^{\alpha}{}_{\mu\nu}&= \delta g^{\alpha\beta}\partial_{\beta}\bar g_{\mu\nu} + \bar g^{\alpha\beta}\partial_{\beta}\delta g_{\mu\nu} \\
    \delta Q_{\mu}{}^{\alpha}{}_{\nu} & = \delta g^{\alpha\beta}\partial_{\mu}\bar g_{\beta\nu} + \bar g^{\alpha\beta}\partial_{\mu}\delta g_{\beta\nu}\\
    \delta Q_{\alpha}{}^{\mu\nu}&= \delta g^{\mu\gamma} \bar g^{\nu\sigma}\bar Q_{\alpha\gamma\sigma} + \bar g^{\mu\gamma}\delta g^{\nu\sigma} \bar Q_{\alpha\gamma\sigma} +\bar g^{\mu\gamma}\bar g^{\nu\sigma}\delta Q_{\alpha\gamma\sigma}
\end{align}

Now we also have some other form : 
\begin{align}
    \delta Q_{\alpha}& = \delta g^{\mu\nu} \partial_{\alpha} \bar g_{\mu\nu} + \bar g^{\mu\nu}\partial_{\alpha}\delta g_{\mu\nu}\\
    \delta Q^{\alpha} &= \delta g^{\mu\nu}\bar g^{\alpha\beta}\partial_{\beta} \bar g_{\mu\nu} + \bar g^{\mu\nu}\delta g^{\alpha\beta}\partial_{\beta}\bar g_{\mu\nu}+ \bar g^{\mu\nu} \bar g^{\alpha\beta}\partial_{\beta}\delta g_{\mu\nu} \\
    \delta \tilde{Q_{\alpha}} &= \delta g^{\mu\nu}\partial_{\mu}\bar g_{\alpha\nu} + \bar g^{\mu\nu}\partial_{\mu} \delta g_{\alpha\nu} \\
    \delta \tilde{ Q^{\alpha}}& = \delta g^{\alpha\beta}\bar g^{\mu\nu}\partial_{\mu}\bar g_{\beta\nu} + \bar g^{\alpha\beta}\delta g^{\mu\nu} \partial_{\mu} \bar g_{\beta\nu} + \bar g^{\alpha\beta}\bar g^{\mu\nu} \partial_{\mu}\delta g_{\beta\nu}
\end{align}

Now we have the disformation tensor and superpotential tensor in perturbative form: 
\begin{align}
    \delta L^{\alpha}{}_{\mu\nu}& = \frac{1}{2}\bigg( \delta Q^{\alpha}{}_{\mu\nu} - \delta Q_{\mu}{}^{\alpha}{}_{\nu} - \delta Q_{\nu}{}^{\alpha}{}_{\mu} \bigg) \\
    \delta P^{\alpha}{}_{\mu\nu}& = -\frac{1}{2}\delta L^{\alpha}{}_{\mu\nu} + \frac{1}{4}\bigg(\delta Q^{\alpha} - \delta \tilde{Q^{\alpha}}\bigg)g_{\mu\nu} + \frac{1}{4}\bigg( Q^{\alpha} -  \tilde{Q^{\alpha}}\bigg)\delta g_{\mu\nu} & \nonumber \\ & -\frac{1}{8}\bigg(\delta^{\alpha}_{\mu
    }\delta Q_{\nu} + \delta^{\alpha}_{\nu}\delta Q_{\mu} \bigg) \nonumber \\
\delta P^{\alpha\mu}_{\phantom{00}\nu}&=\delta g^{\mu\rho}\bar{P}^{\alpha}_{\phantom{00}\rho\nu}+ \bar{g}^{\mu\rho}\delta P^{\alpha}_{\phantom{00}\rho\nu}
\end{align}

Based on the above expressions, we have computed the corresponding components, which are
listed in Table~C. This evaluation uses the scalar perturbations of the metric,
$\delta g_{00} = -2\psi$, $\delta g_{ij} = -2\phi a^{2}\delta_{ij}$ and
$\delta g_{0i} = \delta g_{i0} = 0$, with
$\delta g^{\mu\nu} = -\bar{g}^{\mu\alpha}\bar{g}^{\nu\beta}\delta g_{\alpha\beta}$.
To proceed further, we also need the background quantities
$Q_{\alpha} = g^{\mu\nu}\partial_{\alpha} g_{\nu\mu}$,
$Q^{\alpha} = g^{\mu\nu} g^{\alpha\beta} \partial_{\beta} g_{\mu\nu}$ and
$\tilde{Q}^{\alpha} = g^{\alpha\beta} g^{\mu\nu} \partial_{\mu} g_{\beta\nu}$.

\begin{widetext}

\subsection*{Perturbation Components and Perturbed Field Equation}

\textbf{1. Non-metricity Tensor: $\delta Q^{\alpha}{}_{\mu\nu}$}
\begin{align}
    \delta Q^{0}{}_{00} &= 2\dot\psi \\
    \delta Q^{0}{}_{0i} = \delta Q^{0}{}_{i0} &= 0 \\
    \delta Q^{0}{}_{ij} &= 2\bigg(\dot\phi + 2H(\phi+\psi)\bigg)a^2\delta_{ij} \\
    \delta Q^{i}{}_{00} &= \frac{-2}{a^2}\delta^{ij}\partial_{j}\psi \\
    \delta Q^{i}{}_{0j} = \delta Q^{i}{}_{j0} &= 0 \\
    \delta Q^{i}{}_{jk} &= -2\delta ^{il}\partial_l \phi \delta_{jk} = -2\partial^i \phi \delta_{jk}
\end{align}

\vspace{0.5cm} 

\textbf{2. Non-metricity Tensor (alt trace): $\delta Q_{\mu}{}^{\alpha}{}_{\nu}$}
\begin{align}
    \delta Q_{0}{}^{0}{}_{0} &= 2\dot\psi\\
    \delta Q_{0}{}^{0}{}_{i} &=0\\ \delta Q_{i}{}^{0}{}_{0} &= 2\partial_i\psi\\
    \delta Q_{i}{}^{0}{}_{j} &=0 \\
    \delta Q_{0}{}^{i}{}_{0} &= 0\\
    \delta Q_{0}{}^{i}{}_{j} &= -2\dot\phi \delta^i_j\\\delta Q_{j}{}^{i}{}_{0} &= 0\\
    \delta Q_{j}{}^{i}{}_{k} &= -2\partial_j \phi\delta^i_k
\end{align}

\vspace{0.5cm}

\textbf{3. Disformation tensor: $\delta L^{\alpha}{}_{\mu\nu}$}
\begin{align}
    \delta L^{0}{}_{00} &= -\dot\psi\\
    \delta L^{0}{}_{0i} = \delta L^{0}{}_{i0} &=- \partial_i \psi\\
    \delta L^{0}{}_{ij} &=\bigg(\dot\phi + 2H(\phi+\psi)\bigg)a^2\delta_{ij} \\
    \delta L^{i}{}_{00} &= -\frac{1}{a^2}\delta^{ij}\partial_j \psi\\
    \delta L^{i}{}_{0j} = \delta L^{i}{}_{j0} &=-\dot\phi \delta^i_j \\
    \delta L^{i}{}_{jk} &= -\partial^i\phi\delta_{jk}+ \partial_j\phi\delta^i_k +\partial_k\phi \delta^i_j
\end{align}

\vspace{0.5cm}

\textbf{4. Superpotential tensor: $\delta P^{\alpha}{}_{\mu\nu}$}
\begin{align}
    \delta P^{0}{}_{00} &= 0\\
    \delta P^{0}{}_{0i} = \delta P^{0}{}_{i0} &= \frac{\partial_i(\psi+3\phi)}{4} \\
    \delta P^{0}{}_{ij} &= a^2(\dot\phi+2H(\phi+\psi))\delta_{ij}\\
    \delta P^{i}{}_{00} &= \frac{\delta^{ij}}{a^2}\partial_j\phi \\
    \delta P^{i}{}_{0j} = \delta P^{i}{}_{j0} &= \frac{(\dot\phi-\dot\psi)}{4}\delta^{i}_{j}\\
    \delta P^{i}{}_{jk} &= \frac{1}{2}\bigg(\partial^i\psi\delta_{jk}-\frac{1}{2} \partial_j\psi\delta^i_k -\frac{1}{2}\partial_k\psi \delta^i_j\bigg) -\frac{1}{2}\bigg(\partial^i\phi\delta_{jk}-\frac{1}{2} \partial_j\phi\delta^i_k -\frac{1}{2}\partial_k\phi \delta^i_j\bigg)
\end{align}
\label{tab:perturbation_components_extended}

Recalling equation \eqref{re-field}, we begin to implement the perturbations. We can separate this expansion into three parts as follows:

\begin{align}
    \delta\Bigg(\frac{2}{\sqrt{-g}}\nabla_{\alpha}(f_Q\sqrt{-g}P^{\alpha\mu}_{\phantom{\alpha\mu}\nu})\Bigg)=\delta(\frac{2}{\sqrt{-g}})\partial_{\alpha}(f_Q\sqrt{-g}P^{\alpha\mu}_{\phantom{\alpha\mu}\nu}) + \frac{2}{\sqrt{-g}}\partial_{\alpha}\bigg[(\delta f_Q)\sqrt{-g}P^{\alpha\mu}_{\phantom{\alpha\mu}\nu}+f_Q (\delta\sqrt{-g})P^{\alpha\mu}_{\phantom{\alpha\mu}\nu}+f_Q\sqrt{-g}(\delta P^{\alpha\mu}_{\phantom{\alpha\mu}\nu})\bigg]
\end{align}

Applying the same analogy to the second part, we obtain:

\begin{align}
    \delta \bigg(f_QP^{\mu}_{\phantom{\mu}\alpha\beta} Q_{\nu}^{\alpha\beta}\bigg)= \delta f_Q P^{\mu}_{\phantom{\mu}\alpha\beta} Q_{\nu}^{\alpha\beta}+ f_Q\delta P^{\mu}_{\phantom{\mu}\alpha\beta} Q_{\nu}^{\alpha\beta}+ f_QP^{\mu}_{\phantom{\mu}\alpha\beta} \delta Q_{\nu}^{\alpha\beta}
\end{align}

The remaining term on the left-hand side (LHS) is straightforward to evaluate, yielding $\delta(f\delta^{\mu}_{\nu})= \delta f \delta^{\mu}_{\nu}$. Similarly, we can expand the final term on the right-hand side (RHS) as follows:

\begin{align}
    \delta \Bigg(\frac{f_{\mathcal{L}_m}}{2}(\delta^{\mu}_{\nu}\mathcal{L}_m-T^{\mu}_{\nu})\Bigg) = \frac{1}{2}\delta f_{\mathcal{L}_m}(\delta^{\mu}_{\nu}\mathcal{L}_m-T^{\mu}_{\nu})+ \frac{f_{\mathcal{L}_m}}{2}(\delta^{\mu}_{\nu}\delta \mathcal{L}_m-\delta T^{\mu}_{\nu})
\end{align}

Next, we evaluate this component by component. Here, we focus specifically on the $(^0_0)$ and $(^0_i)$ components (where $i$ denotes the spatial components, $i=x,y,z$), as only these two equations are required to derive the growth structure. 

For the background superpotential components, we have $P^{0}_{\phantom{0}00}=0$ and $\delta P^{00}_{\phantom{00}0}=0$. Similarly, $P^{i0}_{\phantom{00}0}=0$, $P^{00}_{\phantom{00}i}=0$, and finally $P^{j0}_{\phantom{00}i}=-\frac{H}{4}\delta^{j}_{i}$. Consequently, it is clear that $\partial_j P^{j0}_{\phantom{00}i}=0$. Therefore, the coefficient of the $\delta\left(\frac{2}{\sqrt{-g}}\right)$ term will be zero. The variation of the metric determinant measure is given by the standard identity $\delta\sqrt{-g} = \frac{1}{2}\sqrt{-g}g^{\mu\nu}\delta g_{\mu\nu}=\sqrt{-g}(\psi - 3\phi)$. For the perturbation of the non-metricity scalar $\delta Q$, we can adopt the expression from \cite{Xu:2019sbp}, which is given by:

\begin{align}
    \delta Q &= 2 P_{\alpha\nu\rho} \nabla^\alpha \delta g^{\nu\rho} - \left( P_{\mu\alpha\beta} Q_{\nu}^{\ \ \alpha\beta} - 2 Q^{\alpha\beta}_{\phantom{00} \mu} P_{\alpha\beta\nu} \right) \delta g^{\mu\nu}, \nonumber\\
    &= 2 P_{\alpha\nu\rho} \partial^0 \delta g^{ij} - \left( P_{0ij} Q_{0}^{\phantom{0}ij} - 2 Q^{0k}_{\phantom{00}  i} P_{0mj} \right) \delta g^{ij} \nonumber \\
    &= -12H(\dot\phi+H\psi)
\end{align}

In the computation of the above equation, we have considered that $P_{\alpha\nu\rho}= g_{\alpha\beta}P^{\beta}_{\phantom{0}\nu\rho}$ and $Q^{\alpha\beta}_{\phantom{00} \mu}= g^{\alpha\rho}g^{\beta\nu}Q_{\rho\nu\mu}$. The indices $i,j,k,m$ all represent spatial coordinates. 
Thus, we obtain the components for the first step. However, we have the component $\delta P^{i 0}_{\phantom{i0}0}= -\frac{\delta^{ij}}{a^2}\partial_{j}\phi $, which is the only one that exists here:

\begin{align}
    \delta\Bigg(\frac{2}{\sqrt{-g}}\nabla_{\alpha}(f_Q\sqrt{-g}P^{\alpha 0}_{\phantom{\alpha 0}0})\Bigg)&= \frac{2}{\sqrt{-g}}(\sqrt{-g})f_Q\partial_i(\delta P^{i 0}_{\phantom{i0}0})= -\frac{2 f_Q}{a^2}\nabla^2\phi \\
    \delta\Bigg(\frac{2}{\sqrt{-g}}\nabla_{\alpha}(f_Q\sqrt{-g}P^{\alpha 0}_{\phantom{\alpha 0}i})\Bigg)&= \frac{2}{\sqrt{-g}}\partial_0\bigg(\sqrt{-g}f_Q\delta P^{0 0}_{\phantom{00}i}\bigg)+\frac{2}{\sqrt{-g}}\bigg(\sqrt{-g}f_Q\partial_j(\delta P^{j 0}_{\phantom{j0}i}) \bigg) \nonumber \\ 
    &\quad +\frac{2}{\sqrt{-g}}\bigg(\partial_j\bigg(\delta(\sqrt{-g}f_Q)\bigg) P^{j 0}_{\phantom{j0}i} \bigg) \nonumber \\
    &= -f_Q\Bigg[ 2 \partial_i(\dot\phi +H\psi)\left(1+3H^2\frac{f_{QQ}}{f_Q}\right)+6H\dot{H}\frac{f_{QQ}}{f_Q}\partial_i(\psi+3\phi) + 6H\partial_i\phi\Bigg]
\end{align}

Now, evaluating the $(^0_0)$ and $(^0_i)$ components for the second part, we can expand the terms as $P^{0}_{\phantom{\mu}\alpha\beta} Q_{0}^{\alpha\beta}=P^{0}_{\phantom{\mu}00} Q_{0}^{\phantom{0}00}+P^{0}_{\phantom{\mu}0i} Q_{0}^{\phantom{0}0i}+P^{0}_{\phantom{\mu}i0} Q_{0}^{\phantom{0}i0}+P^{0}_{\phantom{\mu}ij} Q_{0}^{\phantom{0}ij}$ and $P^{0}_{\phantom{\mu}\alpha\beta} Q_{i}^{\alpha\beta}=P^{0}_{\phantom{\mu}00} Q_{i}^{\phantom{0}00}+P^{0}_{\phantom{\mu}0j} Q_{i}^{\phantom{0}0j}+P^{0}_{\phantom{\mu}j0} Q_{i}^{\phantom{0}j0}+P^{0}_{\phantom{\mu}jk} Q_{i}^{\phantom{0}jk}$. This allows us to write:

\begin{align}
\delta \bigg(f_QP^{0}_{\phantom{\mu}\alpha\beta} Q_{0}^{\alpha\beta}\bigg)&= \delta f_Q (P^{0}_{\phantom{\mu}ij} Q_{0}^{\phantom{0}ij}) +  f_Q( \delta P^{0}_{\phantom{\mu}ij} Q_{0}^{\phantom{0}ij}) + f_Q(P^{0}_{\phantom{\mu}ij} \delta Q_{0}^{\phantom{0}ij}) \nonumber\\& =12Hf_Q(\dot\phi + H\psi)\left(1+ 6H^2\frac{f_{QQ}}{f_Q}\right) \\
\delta \bigg(f_Q P^{0}_{\phantom{\mu}\alpha\beta} Q_{i}^{\alpha\beta}\bigg)&=  f_Q(P^{0}_{\phantom{\mu}jk} \delta Q_{i}^{\phantom{0}jk}) = 6f_QH\partial_i\phi
\end{align}

We can now also write $\delta f = \frac{\partial f}{\partial Q}\delta Q+\frac{\partial f}{\partial \mathcal{L}_m}\delta \mathcal{L}_m = -12Hf_Q(\dot\phi +H\psi)+ f_{\mathcal{L}_m}\delta \mathcal{L}_m$. The final step we need to perform is:

\begin{align}
\delta \Bigg(\frac{f_{\mathcal{L}_m}}{2}(\delta^{0}_{0}\mathcal{L}_m-T^{0}_{0})\Bigg) &= \frac{1}{2}\frac{\partial f_{\mathcal{L}_m}}{\partial \mathcal{L}_m}(\mathcal{L}_m +\rho)\delta \mathcal{L}_m + \frac{1}{2}f_{\mathcal{L}_m}(\delta \mathcal{L}_m +\delta \rho)\\
\delta \Bigg(\frac{f_{\mathcal{L}_m}}{2}(\delta^{0}_{i}\mathcal{L}_m-T^{0}_{i})\Bigg)&= -\frac{ f_{\mathcal{L}_m}}{2}(\bar\rho+\bar p)av_i
\end{align}

Thus, the first final equation will be:

\begin{align}
    \delta\Bigg(\frac{2}{\sqrt{-g}}\nabla_{\alpha}(f_Q\sqrt{-g}P^{\alpha 0}_{\phantom{\alpha 0}0})\Bigg)+ \delta \bigg(f_QP^{0}_{\phantom{\mu}\alpha\beta} Q_{0}^{\alpha\beta}\bigg) + \frac{1}{2}\delta f \delta^{0}_{0}&= \delta \Bigg(\frac{f_{\mathcal{L}_m}}{2}(\delta^{0}_{0}\mathcal{L}_m-T^{0}_{0})\Bigg) \nonumber \\
    \text{Or, }  -\frac{2 f_Q}{a^2}\nabla^2\phi + 12Hf_Q(\dot\phi + H\psi)\left(1+ 6H^2\frac{f_{QQ}}{f_Q}\right) &\nonumber \\
    -6Hf_Q(\dot\phi +H\psi)+ \frac{1}{2}f_{\mathcal{L}_m}\delta \mathcal{L}_m&=\frac{1}{2}f_{\mathcal{L}_m\mathcal{L}_m}(\bar{\mathcal{L}}_m+\bar \rho)\delta \mathcal{L}_m + \frac{1}{2}f_{\mathcal{L}_m}(\delta \mathcal{L}_m +\delta \rho)
\end{align}

Yielding the final equation:

\begin{align}
    \nabla^2\phi -6a^2H\bigg(\dot\phi +H\psi\bigg)\left(1 + \frac{12H^2f_{QQ}}{f_Q}\right) = -a^2\bigg [\left(\frac{f_{\mathcal{L}_m}}{4f_Q}\right)\delta \rho +\frac{f_{\mathcal{L}_m\mathcal{L}_m}}{4f_Q}\bigg(\bar{\mathcal{L}}_m + \bar\rho\bigg)\delta \mathcal{L}_m \bigg] 
\end{align}

Similarly, for the $(^0_{i})$ component, we obtain the result:

\begin{align}
    \delta\Bigg(\frac{2}{\sqrt{-g}}\nabla_{\alpha}(f_Q\sqrt{-g}P^{\alpha 0}_{\phantom{\alpha 0}i})\Bigg)+ \delta \bigg(f_QP^{0}_{\phantom{\mu}\alpha\beta} Q_{i}^{\phantom{0}\alpha\beta}\bigg) + \frac{1}{2}\delta f \delta^{0}_{i}&= \delta \Bigg(\frac{f_{\mathcal{L}_m}}{2}(\delta^{0}_{i}\mathcal{L}_m-T^{0}_{i})\Bigg) \nonumber \\
    \text{Or, } -f_Q\Bigg[ 2 \partial_i(\dot\phi +H\psi)\left(1+3H^2\frac{f_{QQ}}{f_Q}\right)+6H\dot{H}\frac{f_{QQ}}{f_Q}\partial_i(\psi+3\phi) &+ 6H\partial_i\phi\Bigg]+6Hf_Q\partial_i\phi = -\frac{ f_{\mathcal{L}_m}}{2}(\bar\rho+\bar p)av_i \nonumber \\
    \text{Or, } \partial_i(\dot\phi +H\psi)\left(1+3H^2\frac{f_{QQ}}{f_Q}\right)+3H\dot{H}\frac{f_{QQ}}{f_Q}\partial_i(\psi+3\phi)&=\frac{ f_{\mathcal{L}_m}}{4f_Q}(\bar\rho+\bar p)av_i 
\end{align}

Taking an additional derivative with a contravariant index (such as $\partial^i$) and using the identity $\nabla^2 \equiv \partial ^i\partial_i$ along with our velocity divergence identity in physical cosmic time $\theta \equiv \frac{1}{a}\partial^iv_i$, we find that $\partial^i(av_i)=a^2\theta$. The equation then becomes:

\begin{align}
    \nabla^2\bigg(\dot\phi +H\psi\bigg)\left(1 + \frac{3H^2f_{QQ}}{f_Q}\right) = \left(\frac{f_{\mathcal{L}_m}}{4f_Q}\right) \bigg(\bar\rho + \bar p\bigg)a^2\theta - 3H\dot H \frac{f_{QQ}}{f_Q} \nabla^2\bigg(\psi + 3\phi\bigg)
\end{align}

\subsection*{Perturbed Euler and Continuity Equation}

Integration of the connection field equation ~\eqref{conn} introduces a conserved tensor density. Following the standard covariant formulation, we express this by isolating the pure tensor part of the integration constant, $A^{\nu}_{\phantom{\nu}\alpha}$, from the metric determinant:
\begin{equation}
    \nabla_{\mu} \left( 4\sqrt{-g}f_Q P^{\mu\nu}_{\phantom{\mu\nu}\alpha} + H_{\alpha}^{\phantom{\alpha}\mu\nu} \right) = \sqrt{-g} A^{\nu}_{\phantom{\nu}\alpha} \,.
    \label{eq:connection_integrated}
\end{equation}
Taking the covariant derivative of the right-hand side yields the constraint equation for the auxiliary tensor:
\begin{equation}
    \nabla_{\nu} \left( \sqrt{-g} A^{\nu}_{\phantom{\nu}\alpha} \right) = \sqrt{-g} \nabla_{\nu} A^{\nu}_{\phantom{\nu}\alpha} + \frac{\sqrt{-g}}{2} Q_{\nu} A^{\nu}_{\phantom{\nu}\alpha} = 0 \,.
    \label{eq:A_constraint}
\end{equation}

By rearranging Eq. (\ref{eq:connection_integrated}) and matching the index contractions to our generalized non-conservation equation, we obtain an explicit substitution for the geometric superpotential derivative:
\begin{equation}
    \nabla_{\alpha} \left( f_Q\sqrt{-g} P^{\alpha\mu}_{\phantom{\alpha\mu}\nu} \right) = - \frac{1}{4} \nabla_{\alpha} H^{\alpha\mu}_{\phantom{\alpha\mu}\nu} + \frac{1}{4} \sqrt{-g} A^{\mu}_{\phantom{\mu}\nu} \,.
    \label{eq:superpotential_sub}
\end{equation}
Substituting Eq. (\ref{eq:superpotential_sub}) into the generalized energy-momentum divergence equation yields:
\begin{equation}
    D_{\mu}T_{\nu}^{\mu} = \frac{(\delta_{\nu}^{\mu}\mathcal{L}_m - T_{\nu}^{\mu})\partial_{\mu}f_{\mathcal{L}_m}}{f_{\mathcal{L}_m}} + \frac{1}{f_{\mathcal{L}_m}\sqrt{-g}} \left[ \nabla_{\alpha}\nabla_{\mu}H^{\alpha\mu}_{\phantom{\alpha\mu}\nu} + \frac{1}{2}Q_{\mu}\nabla_{\alpha}H^{\alpha\mu}_{\phantom{\alpha\mu}\nu} - \frac{1}{2}\sqrt{-g}Q_{\mu}A^{\mu}_{\phantom{\mu}\nu} \right] \,.
\end{equation}

Now we can also rewrite the identity of the above equation using expression ~\eqref{eq:A_constraint} like that;

\begin{equation}
    D_{\mu}T_{\nu}^{\mu} = \frac{(\delta_{\nu}^{\mu}\mathcal{L}_m - T_{\nu}^{\mu})\partial_{\mu}f_{\mathcal{L}_m}}{f_{\mathcal{L}_m}} + \frac{1}{f_{\mathcal{L}_m}} \left[ \frac{\nabla_{\alpha}\nabla_{\mu}H^{\alpha\mu}_{\phantom{\alpha\mu}\nu}}{\sqrt{-g}} + \frac{1}{2\sqrt{-g}}Q_{\mu}\nabla_{\alpha}H^{\alpha\mu}_{\phantom{\alpha\mu}\nu} + \nabla_{\mu}A^{\mu}_{\phantom{\mu}\nu} \right] \,.
\end{equation}

As we know $\nabla_\mu(\sqrt{-g})= \frac{1}{2}\sqrt{-g}Q_{\mu}$ so with the help of it we can say $\nabla_\mu(\frac{1}{\sqrt{-g}})= -\frac{Q_\mu}{2\sqrt{-g}}$ and to further proceed for the master equation; we have to do:

\begin{align}
    \nabla_{\mu}\bigg(\frac{1}{\sqrt{-g}}\nabla_{\alpha}H_{\nu}^{\phantom{0}\alpha\mu}\bigg) = \frac{1}{\sqrt{-g}}\nabla_{\mu}\nabla_{\alpha}H_{\nu}^{\phantom{0}\alpha\mu} + \nabla_{\mu}\bigg(\frac{1}{\sqrt{-g}}\bigg)\nabla_{\alpha}H_{\nu}^{\phantom{0}\alpha\mu} = \frac{1}{\sqrt{-g}}\nabla_{\mu}\nabla_{\alpha}H_{\nu}^{\phantom{0}\alpha\mu} -\frac{Q_{\mu}}{2\sqrt{-g}}\nabla_{\alpha}H_{\nu}^{\phantom{00}\alpha\mu}.
\end{align}

So from above equation we can get the extract the $Q_\mu$ linked term like;

\begin{align}
    \frac{Q_{\mu}}{2\sqrt{-g}}\nabla_{\alpha}H_{\nu}^{\phantom{0}\alpha\mu}= \frac{1}{\sqrt{-g}}\nabla_{\mu}\nabla_{\alpha}H_{\nu}^{\phantom{0}\alpha\mu} -  \nabla_{\mu}\bigg(\frac{1}{\sqrt{-g}}\nabla_{\alpha}H_{\nu}^{\phantom{0}\alpha\mu}\bigg)
\end{align}

This yielding the full and final equation:
\begin{align}
    D_{\mu}T_{\nu}^{\mu} = \frac{(\delta_{\nu}^{\mu}\mathcal{L}_m - T_{\nu}^{\mu})\partial_{\mu}f_{\mathcal{L}_m}}{f_{\mathcal{L}_m}} + \frac{1}{f_{\mathcal{L}_m}} \left[ \frac{2\nabla_{\alpha}\nabla_{\mu}H^{\alpha\mu}_{\phantom{\alpha\mu}\nu}}{\sqrt{-g}} - \nabla_{\mu}\bigg(\frac{1}{\sqrt{-g}}\nabla_{\alpha}H_{\nu}^{\phantom{0}\alpha\mu}\bigg) + \nabla_{\mu}A^{\mu}_{\phantom{\mu}\nu} \right] \,.
\end{align}

In applying this framework to a cosmological setting, we adopt the standard homogeneous and isotropic Friedmann-Lema\^itre-Robertson-Walker (FLRW) background. Furthermore, we model the cosmic matter content exclusively as a macroscopic perfect fluid. Consequently, the matter Lagrangian does not account for particles possessing intrinsic spin that couple to the spacetime geometry, such as fermions. In the absence of such microscopic structure, the matter Lagrangian exhibits no functional dependence on the affine connection. Therefore, the hypermomentum tensor inherently vanishes, allowing us to rigorously set $H_{\nu}^{\phantom{\nu}\alpha\mu} = 0$ \cite{BeltranJimenez:2018vdo}.  

So in the absence of hypermomentum $H_{\nu}^{\phantom{\nu}\alpha\mu}$ if  we implement perturbation on the equation it stands for;

\begin{align}
    \delta(D_{\mu}T^{\mu}_{\nu}) = \frac{(\delta_{\nu}^{\mu}\delta \mathcal{L}_m - \delta T_{\nu}^{\mu})\partial_{\mu}f_{\mathcal{L}_m}}{f_{\mathcal{L}_m}}+\frac{(\delta_{\nu}^{\mu}\mathcal{L}_m - T_{\nu}^{\mu})\partial_{\mu} \delta f_{\mathcal{L}_m}}{f_{\mathcal{L}_m}} -\frac{f_{\mathcal{L}_m\mathcal{L}_m}}{f_{\mathcal{L}_m}}\frac{(\delta_{\nu}^{\mu}\mathcal{L}_m - T_{\nu}^{\mu})\partial_{\mu}f_{\mathcal{L}_m}}{f_{\mathcal{L}_m}}\delta \mathcal{L}_m  \nonumber \\ + \frac{1}{f_{\mathcal{L}_m}}\partial_{\mu}(\delta A^{\mu}_{\phantom{\mu}\nu})-\bigg(\frac{f_{\mathcal{L}_m\mathcal{L}_m}}{f_{\mathcal{L}_m}^2}\delta \mathcal{L}_m\bigg)\partial_{\mu} A^{\mu}_{\phantom{\mu}\nu}. 
    \label{pcon}
\end{align}

\subsection*{Case I: $\nu=0$, Perturbed Continuity Equation}
To derive the perturbed continuity equation, we evaluate the general conservation equation for the $\nu=0$ case. In our theoretical framework, the matter Lagrangian is chosen as $\mathcal{L}_m=\rho$. Consequently, under linear perturbations, we have $\delta \mathcal{L}_m= \delta \rho= \bar{\rho}\delta=\bar{L}_{m}\delta$, which yields:

\begin{align}
    \delta(D_{\mu}T^{\mu}_{\nu=0}) = \frac{(\delta \mathcal{L}_m + \delta \rho)\dot f_{\mathcal{L}_m}}{f_{\mathcal{L}_m}}+\frac{(\mathcal{L}_m +\rho)( f_{\mathcal{L}_m\mathcal{L}_m}\dot{\delta \mathcal{L}_m}+\dot{f}_{\mathcal{L}_m\mathcal{L}_m}\delta \mathcal{L}_m)}{f_{\mathcal{L}_m}} -\frac{f_{\mathcal{L}_m\mathcal{L}_m}}{f_{\mathcal{L}_m}}\frac{(\mathcal{L}_m + \rho)\dot{f}_{\mathcal{L}_m}}{f_{\mathcal{L}_m}}\delta \mathcal{L}_m  \nonumber  \\ + \frac{1}{f_{\mathcal{L}_m}}(\partial_{0}(\delta A^{0}_{\phantom{\mu}0})+\partial_{i}(\delta A^{i}_{\phantom{\mu}0}))-\bigg(\frac{f_{\mathcal{L}_m\mathcal{L}_m}}{f_{\mathcal{L}_m}^2}\delta \mathcal{L}_m\bigg)(\partial_{0} A^{0}_{\phantom{\mu}0}+\partial_{i} A^{i}_{\phantom{\mu}0}).
\end{align}

From standard cosmological literature, the covariant derivative $D_{\mu}$ associated with the Levi-Civita connection is defined as:

\begin{equation}
    D_{\mu}T^{\mu}_{\nu}= \partial_{\mu}T^{\mu}_{\nu} + \begin{Bmatrix} \mu \\ \mu \ \rho \end{Bmatrix} T^{\rho}_{\nu}-  \begin{Bmatrix} \rho \\ \mu \ \nu \end{Bmatrix}T^{\mu}_{\rho}
\end{equation}

Applying linear perturbations to this expression leads to:

\begin{equation}
    \delta \bigg(D_{\mu}T^{\mu}_{\nu}\bigg)= \partial_{\mu}(\delta T^{\mu}_{\nu} ) + \delta\begin{Bmatrix} \mu \\ \mu \ \rho \end{Bmatrix} T^{\rho}_{\nu}+\begin{Bmatrix} \mu \\ \mu \ \rho \end{Bmatrix} \delta T^{\rho}_{\nu}- \delta \begin{Bmatrix} \rho \\ \mu \ \nu \end{Bmatrix}T^{\mu}_{\rho} - \begin{Bmatrix} \rho \\ \mu \ \nu \end{Bmatrix}\delta T^{\mu}_{\rho}
\end{equation}

Here, $\begin{Bmatrix} \rho \\ \mu \ \nu \end{Bmatrix}$ denotes the Christoffel symbol within the Levi-Civita framework, defined by the standard relation $\begin{Bmatrix} \rho \\ \mu \ \nu \end{Bmatrix} = \frac{1}{2}g^{\rho\lambda}\bigg(\partial_{\nu}g_{\lambda\mu}+\partial_{\mu}g_{\lambda\nu}-\partial_{\lambda}g_{\mu\nu}\bigg)$. The corresponding perturbed Christoffel symbol evaluates to:

\begin{align}
    \delta \begin{Bmatrix} \rho \\ \mu \ \nu \end{Bmatrix} = \frac{1}{2}\delta g^{\rho\lambda}\bigg(\partial_{\nu}g_{\lambda\mu}+\partial_{\mu}g_{\lambda\nu}-\partial_{\lambda}g_{\mu\nu}\bigg)+ \frac{1}{2}g^{\rho\lambda}\bigg(\partial_{\nu}\delta g_{\lambda\mu}+\partial_{\mu} \delta g_{\lambda\nu}-\partial_{\lambda}\delta g_{\mu\nu}\bigg)
\end{align}

Setting $\nu=0$, we expand the perturbed covariant derivative as:

\begin{align}
    \delta(D_{\mu}T^{\mu}_{\nu=0})= \partial_{0}\delta T^{0}_{0} +\partial_{j}\delta T^{j}_{0}+ \bigg(\delta\begin{Bmatrix} 0 \\ 0 \ 0 \end{Bmatrix}  + \delta\begin{Bmatrix} j \\ j \ 0 \end{Bmatrix} \bigg)T^{0}_{0}+\bigg(\begin{Bmatrix} 0 \\ 0 \ 0 \end{Bmatrix}+ \begin{Bmatrix} j \\ j \ 0 \end{Bmatrix}\bigg )\delta T^{0}_{0}+\bigg(\begin{Bmatrix} 0 \\ 0 \ i \end{Bmatrix}+ \begin{Bmatrix} j \\ j \ i \end{Bmatrix}\bigg )\delta T^{i}_{0} \nonumber \\ -\delta\begin{Bmatrix} 0 \\ 0 \ 0 \end{Bmatrix}  T^{0}_{0}- \delta\begin{Bmatrix} j \\ i \ 0 \end{Bmatrix}  T^{i}_{j} -\begin{Bmatrix} 0 \\ 0 \ 0 \end{Bmatrix}\delta T^{0}_{0}-\begin{Bmatrix} j \\ 0 \ 0 \end{Bmatrix}\delta T^{0}_{j}-\begin{Bmatrix} 0 \\ j \ 0 \end{Bmatrix}\delta T^{j}_{0}-\begin{Bmatrix} j \\ i \ 0 \end{Bmatrix}\delta T^{i}_{j}
\end{align}

Considering a pressureless dust matter scenario ($p=0$, $\delta p=0$), the speed of sound satisfies $c_{S}^2=\frac{\delta p}{\delta \rho}=0$ and the background equation of state parameter is $w =\frac{\bar p}{\bar\rho}=0$. Utilizing the background continuity equation~\eqref{bcon}, the geometric side reduces to:

\begin{align}
    \delta(D_{\mu}T^{\mu}_{\nu=0})&= -\bigg(\dot{\delta \rho}+3H(\delta\rho+\delta p)+({\bar{{\rho}}}+{\bar{p}})(\theta -3\dot\phi)\bigg) \nonumber \\ &=-\bar{\rho}\bigg(\dot{\delta}- \frac{(\bar{\mathcal{L}_m}+\bar{\rho})}{\bar{\rho}}\frac{\dot{f}_{\mathcal{L}_m}}{f_{\mathcal{L}_m}}\delta+(\theta-3\dot{\phi})\bigg)
\end{align}

Equating the perturbed geometric expansion to the perturbed source terms allows us to write the relation as:

\begin{align}
\dot{\delta}- \frac{(\bar{L}_{m}+\bar{\rho})}{\bar{\rho}}\frac{\dot{f}_{\mathcal{L}_m}}{f_{\mathcal{L}_m}}\delta+\theta= -\frac{(\bar{L}_{m}+\bar{\rho})}{\bar{\rho}}\Bigg[\frac{\dot{f}_{\mathcal{L}_m}}{f_{\mathcal{L}_m}}\delta +\delta\bigg(\bar{L}_m\bigg(\frac{\dot{f}_{{\mathcal{L}_m\mathcal{L}_m}}}{f_{\mathcal{L}_m}}-\frac{f_{\mathcal{L}_m\mathcal{L}_m}}{f_{\mathcal{L}_m}}\frac{\dot{f}_{\mathcal{L}_m}}{f_{\mathcal{L}_m}}\bigg)+\dot{L}_{m}\bigg(\frac{f_{\mathcal{L}_m\mathcal{L}_m}}{f_{\mathcal{L}_m}}\bigg)\bigg) +\dot{\delta}\bigg(\bar{L}_m\frac{f_{\mathcal{L}_m\mathcal{L}_m}}{f_{\mathcal{L}_m}}\bigg)\Bigg]
\end{align}

This expression can be simplified further into the following form:

\begin{align}
    \dot{\delta}+\theta = -\frac{(\bar{L}_{m}+\bar{\rho})}{\bar{\rho}}\Bigg(
    \frac{d}{dt}\bigg(\frac{f_{\mathcal{L}_m\mathcal{L}_m}}{f_{\mathcal{L}_m}}\mathcal{L}_m\bigg)\delta +\dot{\delta} \bigg(\frac{f_{\mathcal{L}_m\mathcal{L}_m}}{f_{\mathcal{L}_m}}\mathcal{L}_m\bigg) \Bigg)
\end{align}

By imposing the background condition $\bar{L}_m=\bar\rho$ alongside our specific functional choice $f_{\mathcal{L}_m}=\alpha \nu \mathcal{L}_m^{\nu-1}$, the coupled functional combination reduces to a constant value, namely $\frac{f_{\mathcal{L}_m\mathcal{L}_m}}{f_{\mathcal{L}_m}}\mathcal{L}_m = (\nu-1)$. Consequently, the final perturbed continuity equation simplifies to:

\begin{align}
    (2\nu-1)\dot{\delta}+\theta=0
\end{align}

\subsection*{Case II: $\nu=i$, Perturbed Euler Equation}

To derive the perturbed Euler equation, we focus on the spatial components by setting the index to $\nu=i$, where $i$ denotes the spatial coordinates ($i=x,y,z$). The geometric perturbation expansion yields:

\begin{align}
   \delta( D_{\mu}T^{\mu}_{\nu=i}) = \partial_0\delta T^{0}_{i} + \partial_{j}\delta T^{j}_{i} +\bigg(\delta\begin{Bmatrix} 0 \\ 0 \ j \end{Bmatrix}  + \delta\begin{Bmatrix} k \\ k \ j \end{Bmatrix} \bigg)T^{j}_{i}+\bigg(\begin{Bmatrix} 0 \\ 0 \ 0 \end{Bmatrix}+ \begin{Bmatrix} j \\ j \ 0 \end{Bmatrix}\bigg )\delta T^{0}_{i}+\bigg(\begin{Bmatrix} 0 \\ 0 \ j \end{Bmatrix}+ \begin{Bmatrix} k \\ k \ j \end{Bmatrix}\bigg )\delta T^{j}_{i} \nonumber \\ -\delta\begin{Bmatrix} 0 \\ 0 \ i \end{Bmatrix}  T^{0}_{0}- \delta\begin{Bmatrix} j \\ k \ i \end{Bmatrix}  T^{k}_{j} -\begin{Bmatrix} 0 \\ 0 \ i \end{Bmatrix}\delta T^{0}_{0}-\begin{Bmatrix} j \\ 0 \ i \end{Bmatrix}\delta T^{0}_{j}-\begin{Bmatrix} 0 \\ j \ i \end{Bmatrix}\delta T^{j}_{0}-\begin{Bmatrix} j \\ k \ i \end{Bmatrix}\delta T^{k}_{j}
\end{align}

Evaluating all the background and perturbed Christoffel symbols, while continuing to enforce the dust conditions $\bar p=0$ and $\delta p=0$, we obtain:

\begin{align}
    \delta( D_{\mu}T^{\mu}_{\nu=i})= \frac{d}{dt}\bigg((\bar{\rho}+\bar{p})av_i\bigg) + (\bar\rho+\bar p)\partial_{i}\psi + 3H(\bar\rho+\bar p)av_i \nonumber \\
    =a\bar \rho(\dot{v_i}+ Hv_i + \frac{\dot{\rho}}{\rho}v_i + 3Hv_i +\frac{1}{a}\partial_{i}\psi)
\end{align}

Recalling the general perturbation formula \eqref{pcon} for the $\nu=i$ component, the full governing equation is written as:

\begin{align}
    a\bar \rho(\dot{v_i}+ Hv_i + \frac{\dot{\rho}}{\rho}v_i + 3Hv_i +\frac{1}{a}\partial_{i}\psi) = - (\bar\rho a v_i)\frac{\dot{f}_{\mathcal{L}_m}}{f_{\mathcal{L}_m}} + \delta^{j}_{i} \bar{L}^2_m \frac{f_{\mathcal{L}_m\mathcal{L}_m}}{f_{\mathcal{L}_m}}\partial_{j}\delta
\end{align}

Substituting the background continuity equation under the dust condition $\bar{p}=0$, which takes the form $\dot\rho+3H\rho=-(\mathcal{L}_m+\rho)\frac{\dot{f}_{\mathcal{L}_m}}{f_{\mathcal{L}_m}}$, we rearrange the expression as follows:

\begin{align}
    a\bar \rho(\dot{v_i}+ Hv_i + (-\frac{(\bar{\mathcal{L}}_m+\bar\rho)}{\bar\rho}\frac{\dot{f}_{\mathcal{L}_m}}{f_{\mathcal{L}_m}})v_i +\frac{\dot{f}_{\mathcal{L}_m}}{f_{\mathcal{L}_m}}v_i+\frac{1}{a}\partial_{i}\psi) =   \delta^{j}_{i} \bar{L}^2_m \frac{f_{\mathcal{L}_m\mathcal{L}_m}}{f_{\mathcal{L}_m}}\partial_{j}\delta
\end{align}

Applying minor algebraic manipulations and acting on the equation with the spatial divergence operator $\frac{1}{a}\partial^i$, we obtain:

\begin{align}
    \frac{1}{a}\partial^{i}\dot v_i + (H- \frac{\bar{\mathcal{ {L}}}_m}{\bar\rho}\frac{\dot{f}_{\mathcal{L}_m}}{f_{\mathcal{L}_m}})\frac{1}{a}\partial^{i}v_i + \frac{1}{a^2}\nabla^2\psi =  \frac{\bar{\mathcal{L}}^2_m }{\bar\rho}\frac{f_{\mathcal{L}_m\mathcal{L}_m}}{f_{\mathcal{L}_m}}\delta^{j}_{i}\frac{1}{a^2}\partial^i\partial_{j}\delta
\end{align}

We now introduce the standard definition for the velocity divergence, $\theta = \frac{1}{a}\partial^iv_i$, and identify the spatial Laplacian operator via $\delta^{j}_{i}\frac{1}{a^2}\partial^i\partial_{j}\equiv \partial^j\partial_j\equiv\nabla^2$. Accounting for the time derivative of the divergence, $\dot\theta= \frac{1}{a}\partial^i\dot v_i-\frac{\dot a}{a^2}\partial^iv_i= \frac{1}{a}\partial^i\dot v_i-H\theta$, and implementing the identification $\bar {\mathcal{L}}_m=\bar\rho$, we arrive at the final configuration space equation:

 \begin{align}
     \dot\theta+ (2H-\frac{\dot{f}_{\mathcal{L}_m}}{f_{\mathcal{L}_m}})\theta + \frac{1}{a^2}\nabla^2\psi = {\bar{\mathcal{L}}_m }\frac{f_{\mathcal{L}_m\mathcal{L}_m}}{f_{\mathcal{L}_m}}\frac{1}{a^2}\nabla^2\delta
 \end{align}

Finally, performing a Fourier transformation by substituting the spatial operator with its momentum space counterpart, $\nabla^2\rightarrow(-k^2)$, the closed-form perturbed Euler equation in Fourier space reads:

\begin{align}
     \dot\theta+ \bigg(2H-\frac{\dot{f}_{\mathcal{L}_m}}{f_{\mathcal{L}_m}}\bigg)\theta - \frac{1}{a^2}k^2\psi + (\nu-1)\frac{1}{a^2}k^2\delta =0
\end{align}

\end{widetext}

\bibliography{references}

\end{document}